\documentclass[screen,authorversion, nonacm]{acmart}
\usepackage{enumitem}
\usepackage{subcaption}
\usepackage{array}
\AtBeginDocument{%
  }

\begin{document}

\title[Give Me a Choice]{Give Me a Choice: The Consequences of Restricting Choices Through AI-Support for Perceived Autonomy, Motivational Variables, and Decision Performance}
\author{Cedric Faas}
\orcid{0009-0000-7918-4233}
\email{s8cefaas@uni-saarland.de}
\affiliation{
  \institution{Saarland Informatics Campus, Saarland University}
  \city{Saarbrücken}
  \country{Germany}
}
\author{Richard Bergs}
\orcid{0000-0001-9697-2311}
\email{richard.bergs@psychologie.uni-freiburg.de}
\affiliation{
  \institution{Department of Psychology, University of Freiburg}
  \city{Freiburg im Breisgau}
  \country{Germany}
}
\author{Sarah Sterz}
\orcid{0000-0002-5365-2198}
\email{sterz@depend.uni-saarland.de}
\affiliation{
  \institution{Saarland Informatics Campus, Saarland University}
  \city{Saarbrücken}
  \country{Germany}
}
\author{Markus Langer}
\orcid{0000-0002-8165-1803}
\authornote{Both authors supervised this research equally.}
\email{markus.langer@psychologie.uni-freiburg.de}
\affiliation{
  \institution{Department of Psychology, University of Freiburg}
  \city{Freiburg im Breisgau}
  \country{Germany}
}
\author{Anna Maria Feit}
\orcid{0000-0003-4168-6099}
\authornotemark[1]
\email{feit@cs.uni-saarland.de}
\affiliation{
  \institution{Saarland Informatics Campus, Saarland University}
  \city{Saarbrücken}
  \country{Germany}
}


\begin{abstract}
Design optimizations in human-AI collaboration often focus on cognitive aspects like attention and task load. Drawing on work design literature, we propose that effective human-AI collaboration requires broader consideration of human needs (e.g., autonomy) that affect motivational variables (e.g., meaningfulness). In a simulated drone oversight experiment, participants (\textit{N}=274, between-subject) faced 10 critical decision-making scenarios with varying levels of choice restrictions with an AI recommending only 1, 2, 4 or all 6 possible actions. Restricting participants to one selectable action improved task performance (with a perfect AI) but significantly reduced perceived autonomy and work meaningfulness, and these effects intensified over time. In conditions with multiple action choices, participants with higher perceived autonomy performed better. The findings underscore the importance of considering motivational factors to design successful long-term human-AI collaboration at work.

\end{abstract}




\maketitle

\section{Introduction}

AI-based systems can support users in complex or time-sensitive scenarios by providing for example decision recommendations~\cite{bansal_does_2021, bucinca_proxy_2020, green_principles_2019, poursabzi-sangdeh_manipulating_2021, zhang_effect_2020}. Typically, the goal of such systems is to improve overall task performance (e.g., decision accuracy, decision time). Thus, most research has focused on optimizing the accuracy of AI methods~\cite{lu_does_2024, ma_are_2024, meske_investigating_2024, pareek_trust_2024, robbemond_understanding_2022} or the joint human-AI performance, by considering cognitive factors such as users' attention~\cite{Das_2024, Parasuraman_2010}, cognitive load~\cite{Feit2020, herm_impact_2023, hudon_explainable_2021}, or their understanding of AI recommendations~\cite{bansal_does_2021, Langer_2022_trust, Bahner_2008}.


However, there is a history of research in work psychology showing that effectiveness in work tasks, including decision-making, depends not only on human cognition but also on human motivation \cite{campbell1993theory, Hackman_oldham_JDS_1975, Parker_SMART_2023}, which are closely linked to considering psychological needs (e.g., a need for autonomy, a need for competence; \cite{Ryan_Deci_SDT_2000}).
Regarding the design of AI-based systems to support humans at work,  we see only few works so far that consider the influence of motivational aspects on worker satisfaction and performance~\cite{buccinca2024towards, bucinca2024_intrinsicmotivation, de_vreede_design_2021}. Research on human-AI interactions still mostly emphasizes cognitive and performance-related aspects. 
Problematically, when focusing predominantly on optimizing for cognitive aspects, this can even undermine motivational aspects of work. For example, in an attempt to reduce users' cognitive load, designers may consider implementing AI-based decision support that reduces the choices that users have for actions. Simultaneously, such a restriction of choices can reduce users' perceived autonomy potentially undermining people's perceived meaningfulness of their task with meaningfulness being one central variable related to long-term job motivation and satisfaction~\cite{Parker_SMART_2023}. Ironically, this can then also negatively impact long-term human-AI joint performance given the importance of motivation and satisfaction for job performance~\cite{campbell1993theory}.
Thus, we argue that human-AI interaction also needs to be optimized considering motivational aspects.

Motivated by theories from the work design literature (e.g., \cite{Hackman_oldham_JDS_1975,Morgeson_2003,Parker_SMART_2023}), this paper aims to establish a relationship between psychological needs,  motivational variables and users' performance when interacting with AI systems at work. 
Specifically, our goal is to explore how the restriction of choices through AI systems affects the psychological need for autonomy, and how this relates to users' motivation and task performance in AI-supported decision-making tasks. 
We present results from an experimental study that tested the consequences of reducing people's choices for actions in a simulated drone oversight task. Participants were in the role of a human overseer of an autonomous drone that faced different critical situations that required the human overseer to decide for an action on how the drone can safely resolve the situation. 
The four between-participant experimental conditions were that the human overseer received decision support by an AI-based system that offered six actions to resolve the situation, or that of those \emph{Six} actions only \emph{Four} vs. \emph{Two} vs. \emph{One} were selectable while the other options were grayed out (i.e. not recommended by the AI). Following theories in work design, we hypothesized that restricting the number of \textit{Selectable Actions} would affect participants' perceived autonomy.  
We expected that this will impact downstream motivational aspects of work such as perceived meaningfulness, with potential effects on human task performance in the oversight task.

Our findings revealed that having only \emph{One} \textit{Selectable Action}  decreased participants perceived autonomy and perceived meaningfulness of the task compared to being able to choose between \emph{Two} or more actions (while increasing the decision accuracy, since the remaining action was always correct). 
 This effect increased over time. 
 In the conditions that had more than only \emph{One} \textit{Selectable Action}, participants who felt more autonomous were also the ones who showed a better decision accuracy. 

We see three main contributions of this paper. 
First, we highlight the importance of considering motivational aspects in designing beneficial and effective human-AI interactions and highlight the value of the work design literature for this endeavor. 
Second, we show that whereas restricting participants' choices during the oversight task increased their task performance (an effect that strongly depends on the accuracy of the AI-based decision support), this can also reduce their perceived autonomy and perceived meaningfulness of their work. These effects on such motivational aspects of work grew stronger over time showing the need to examine human-AI interactions in the longer term. 
Third, our findings emphasize the importance of evaluating the effects of motivational aspects on joint human-AI performance at work given that participants who felt more autonomous also performed better than those who felt less autonomous.

\section{Related Work}

\subsection{AI-supported decision-making}
Research on AI-supported decision-making covers many domains \cite{Lai_2023}. For example, AI-based systems can support decision-making in domains such as education \cite{rastogi_deciding_2022}, finance \cite{cau_supporting_2023, appelganc_how_2022, alufaisan_does_2021, zhang_effect_2020}, healthcare \cite{appelganc_how_2022, cabitza_painting_2023, calisto_breastscreening-ai_2022, gu_improving_2023, jacobs_how_2021}, hiring \cite{Langer_2022_trust, Lee_2018, Marcinkowski_2020} or law \cite{grgic-hlaca_human_2019, liu_understanding_2021, kahr_it_2023, wang_are_2021}. 

One key goal of AI-supported decision-making is to foster effective decisions. The effectiveness of human-AI collaborations can be measured, for instance, by the joint decision accuracy of humans and AI systems \cite{lu_does_2024, ma_are_2024, meske_investigating_2024, pareek_trust_2024, robbemond_understanding_2022, Schoeffer_2024, swaroop_accuracy-time_2024}. Other measures of effectiveness include the time required for decision-making with AI-based systems, especially in time-critical decision contexts \cite{lu_does_2024, meske_investigating_2024, rastogi_deciding_2022, swaroop_accuracy-time_2024}. In theory, effective human-AI collaboration is possible, and studies show that AI-based systems can improve decision-making across various contexts \cite{Lai_2023}. 

However, recent studies indicate that human-AI collaborations can also lead to suboptimal outcomes, as the joint performance of humans and systems may often be worse than fully automating tasks with an AI-based system \cite{bansal_does_2021, bucinca_proxy_2020, green_principles_2019, poursabzi-sangdeh_manipulating_2021, zhang_effect_2020}. For example, users of AI-based systems may find it difficult to discern accurate from inaccurate system outputs \cite{Green_2022, grgic-hlaca_human_2019, Zerilli_2019}, and, as a consequence, they may override actually accurate system outputs and follow inaccurate ones instead \cite{Green_2019_disparate}. 

Thus, research aims to make human-AI collaborations more effective.  
A large share of attempts to optimize human-AI decision-making to date has been focused on cognitive aspects relating to human decision-making performance. For example, the area of XAI tries to achieve better joint performance by enhancing people's understanding of AI-based decision processes and outputs \cite{Langer_2021_whatdowewant, hoffman_metrics_2019, lima_human_2021, robbemond_understanding_2022, wang_watch_2023}. This should help users decide when to follow, reject, or adapt system outputs \cite{bucinca_trust_2021, Schoeffer_2024}. Other research aims to improve joint decision-making by designing systems that can communicate uncertainty in their outputs, ultimately targeting a similar goal as explanations: providing users with better insights into when to follow or reject system outputs \cite{bansal_does_2021}. In an attempt to encourage more thoughtful decision-making, further research suggests using cognitive forcing methods, such as altering the timing of when users receive AI-based outputs (e.g., only after users have made an initial decision \cite{bucinca_trust_2021, Langer_2020_changing}), or introducing a time lag until people receive the AI-based output \cite{buccinca2024towards}. Other research focuses on optimizing users' cognitive load, for example, by automatically adapting interfaces to their current workload~\cite{Lindlbauer2019} or reducing the perceptual load of users by changing the presentation of information-dense interfaces~\cite{Feit2020}.

While optimizing cognitive aspects in human-AI decision-making is crucial, work psychology literature reviewed in the next section emphasizes the role of psychological needs and motivational factors for worker effectiveness and satisfaction. In line with emerging research highlighting the importance of considering motivational aspects for optimizing human-AI interaction \cite{bucinca2024_intrinsicmotivation, steyvers_three_nodate}, in this paper we want to explore in how far this generalizes to work supported by AI-based systems.

\subsection{Motivational Aspects of human-AI Collaboration at Work}

Work psychology literature emphasizes that effectiveness in work tasks, such as decision-making, depends not only on human cognition but also on human motivation \cite{Hackman_oldham_JDS_1975, Morgeson_2003, Parker_SMART_2023, campbell1993theory}. 
In human-AI interaction, these motivational aspects have only recently begun to receive attention \cite{bucinca2024_intrinsicmotivation, steyvers_three_nodate} with researchers such as Steyvers and Kumar~\cite{steyvers_three_nodate} calling for studies to explore the impact of motivational factors, particularly on the long-term use of AI tools. Therefore, in the following, we focus on work psychology literature to understand the relationship between psychological needs, motivation, and performance, which has a long history of research in the work context (e.g., \cite{Hackman_oldham_JDS_1975}).

The importance of motivational aspects of work can be understood by considering the core functions of human motivation: Motivation gives direction, that is which behavior one chooses to perform, it affects the degree of effort one invests in a behavior, and it impacts the persistence and duration of that effort \cite{Vollmeyer_motivation_2000, campbell1993theory}. All else being equal, a motivated worker should therefore perform better for a longer period of time than a less motivated worker. Furthermore, research typically expects and finds a strong relation between motivation and satisfaction~\cite{Humphrey_2007, Morgeson_2003, Parker_SMART_2023}, meaning that motivated workers are also expected to be more satisfied with their work for a longer period of time.

There are various streams of research that help us understand which factors in human-AI collaboration might contribute to motivational aspects of work. These streams share a common focus on describing \emph{psychological needs} and \emph{psychological processes} that contribute to motivating work environments. 

\emph{Self-Determination Theory} is one of the broadest and most comprehensive theories describing three basic psychological needs that contribute to human motivation: a need for autonomy, competence, and relatedness (\cite{Ryan_Deci_SDT_2000}). The need for autonomy refers to having choices and being the initiator of one's own actions; the need for competence pertains to engaging in optimally challenging tasks and effectively achieving desired outcomes; the need for relatedness refers to feeling a sense of belonging and building relationships with others. 
In the workplace, autonomy is positively associated with motivation and well-being  \cite{Humphrey_2007, Van_den_Broeck_Capturing_2010}. 
Autonomy can also positively affect work performance and effort \cite{Van_den_Broeck_Review_2016, Cerasoli_Meta_2016}. Beyond the work context, research shows positive effects of autonomy on positive affect and well-being \cite{ryan_deci_SDT_2017, Stanley_meta_2021, Tang_review_2019}. 

Most of what is proposed by the Self-Determination Theory is consistent with the key propositions in the work design literature. While there are many theories and models to draw from (see, e.g., \cite{Hackman_oldham_JDS_1975, Hackman_redesign_1980, Morgeson_2003, Morgeson_WDQ_2006}), we decided to build on the SMART model by Parker \& Knight  \cite{Parker_SMART_2023} to further understand how fulfilling psychological needs can contribute to effective human-AI collaborations.  
The \emph{SMART model} refers to five different characteristics of work that influence work satisfaction through different psychological processes: \emph{S}timulating work characteristics, \emph{M}astery of work characteristics, \emph{A}utonomous work characteristics, \emph{R}elational work characteristics, and \emph{T}olerable work characteristics. 
In this paper we particularly focus on the effects of autonomous work characteristics that are supposed to affect users primarily through motivational aspects. Specifically, Parker \& Knight argue that having choices and control with respect to the decisions one makes at work and with respect to how to do their work will lead to a higher perceived autonomy and, consequently, a higher perceived work meaningfulness~\cite{Parker_SMART_2023}. In the long term, this is expected to contribute to a higher motivation at work and to a higher job satisfaction. These propositions are supported by meta-analyses of work psychology studies showing that autonomy relates to variables such as meaningfulness, motivation, and job satisfaction \cite{Humphrey_2007}.

Interdisciplinary research on the effects of implementing AI at work implies that AI-based systems may strongly affect people's autonomy \cite{Parent_Rocheleau_2022, Parker_2020_algorithms}. For example, research on the algorithmic management of Uber drivers and platform workers shows that AI-based tools managing workers can undermine workers' perceived feelings of being in control of how they can perform their work \cite{LangerLanders_2021, moehlmann_2021}. Ulfert et al. found that a system's level of automation influenced users' intention to use the system, but did not find evidence that high-autonomy decision support systems increase stress \cite{ulfert_agent_autonomy}.  Findings from a study by Passalacqua et al. indicate that fully automating the selection of optimal decisions, compared to involving users in the decision selection, negatively impacted their perceived autonomy and motivation during the training for an assembly line task \cite{passalacqua_practice_2024}. In a case study by \citet{Strich_2021}, loan officers, whose primary task was to input information into an AI-based system and communicate the AI-based decision to loan applicants, were found to start manipulating the input data to be able to affect the AI-based outputs and thus regain their autonomy over the AI-controlled decision processes.
\citet{de_vreede_design_2021} conducted a study where 
participants with always available AI-support felt less autonomous but performed better than those who could choose whether the AI-support was revealed for each decision. 
We suspect that the effect on decision accuracy in the experiment by De Vreede et al. was due to the experimental conditions rather than the participants' perceived autonomy. Specifically, the experimental conditions were designed in a way that could make deciding correctly easier if the AI-support was always available. In the condition where participants had to turn on the AI-support, decisions might have been worse because the (mostly accurate) support was used less. This highlights the importance of analyzing the effects of perceived autonomy on decision accuracy while properly accounting for potential effects of the accuracy of available AI-support.

\section{Method}

Motivated by the work design literature, the goal of our research was to explore which role psychological needs and motivational processes play in the design and use of AI-decision support systems at work. Given the research in this field discussed above, we were particularly interested in perceived autonomy as one of three major psychological needs, and how a restriction in users' perceived autonomy might impact motivational aspects such as perceived meaningfulness and in conclusion users' task performance. 

Therefore, we devised an online experiment where participants were tasked to monitor an autonomously flying delivery drone and its sensor values and quickly decide on suitable actions when the drone encountered critical situations (e.g., low battery or strong winds). 

We chose this task because it represented a realistic scenario for decision-making under time constraints that participants would be able to relate to. At the same time it allowed us to  measure the task performance as a result of the drone crashing or being saved.
To integrate AI recommendations and simplify the task for participants inexperienced in drone flying, the monitoring interface provided users with six actions to choose from for how to resolve the critical situation (e.g. landing the drone or flying higher) rather than giving participants full control over the drone's flight. This enabled us to easily integrate different levels of (simulated) AI support. In three of the four experimental conditions, the AI-based system provided additional decision support by graying out certain actions that the AI would not recommend. Given that the grayed-out actions were also not selectable anymore, this restricted participants in their choice and thus their autonomy. 

In the following, we describe our research questions, study design, and data collection. Our study was approved by the department's ethical review board and preregistered via the \hyperlink{https://osf.io/r2jhw/?view_only=af2ff5f27f2b44129961a69baa17df0a}{Open Science Foundation}, where we also published the collected data.

\subsection{Research Questions}
Based on our theoretical reasoning informed by the Self-Determination Theory and the work design literature, we expect that the degree of restriction people experience through AI-support in a decision-making task will affect users in multiple ways. First, we expect that restricting users' choices will negatively affect users' perceived autonomy and will detrimentally affect downstream motivational effects on users’ perceived meaningfulness, task motivation, and task satisfaction. These motivational effects may also negatively impact users' overall task performance. 

However, decreasing the number of \textit{Selectable Actions} in a decision-making task can also reduce users' task load \cite{chernev_choice_2015, fandel_information_2001}. Additionally, in case of a reliable and highly accurate AI-based systems, having few \textit{Selectable Actions} can also directly contribute to increased joint human-AI decision accuracy given that actions that would lead to unfavorable outcomes will not be recommended by the system (see also \cite{de_vreede_design_2021}). Therefore, reducing the number of \textit{Selectable Actions} might positively affect users' decision accuracy and reduce the task load they experience. 
 
We thus face contradicting possible effects of restricting users' choices in the decision making task: restricting choices may reduce users' perceived autonomy which can detrimentally affect motivational aspects, thus also potentially decreasing task performance; but restricting choices may also reduce users' task load or may have a direct positive influence on task performance by reducing the number of ineffective available choices. Given these contradicting effects that could play out differently in our experimental conditions, we refrained from proposing directed hypotheses for our experimental conditions, and instead ask the following research questions: 

Does the difference in the number of \textit{Selectable Actions} affect \ldots
\begin{enumerate}[label={RQ1\alph*.}]
    \item \ldots users’ perceived autonomy?
    \item \ldots users’ perceived meaningfulness?
    \item \ldots users’ task motivation?
    \item \ldots users’ task load?
    \item \ldots the accuracy of decisions made in the drone handover task?
    \item \ldots users’ decision time?
    \item \ldots users’ task satisfaction?
\end{enumerate}

Additionally, we explore whether perceived autonomy, meaningfulness, task motivation, and task satisfaction 
are related to \ldots
\begin{enumerate}[label={RQ2\alph*.}]
    \item \ldots the accuracy of decisions made in the drone handover task.
    \item \ldots users’ decision time.
\end{enumerate}


\subsection{Participants}
We conducted an a-priori calculation of the required sample size with G*Power \cite{faul_gpower_2007}. We determined a required sample size of \textit{N}~=~271, to achieve a power of 1-BETA~=~.95 given an ALPHA level of .05 and expecting a medium effect size of f~=~.25 in a between-participants ANOVA with four conditions.\footnote{Note that consistent with our preregistration, we finally decided to use linear mixed models for most of the analyses given that they can efficiently analyze and communicate the direction of main effects and interaction effects of independent variables. Given the similarity between ANOVA analyses and regression analyses such as linear mixed models, these calculations should also be suited to approximate the required number of participants for linear mixed models.} We decided to conduct the study as an online experiment on Prolific, where we restricted the sampling to English-speaking participants, participants with a prior approval rate greater than 95\%, and who joined Prolific more than a year ago. We collected data from 280 participants (151 female, 129 male, age:~18-71, \textit{M}~=~33.25, \textit{SD}~=~9.81). The participants were compensated with £6.32 for participating in the study (median completion time: 40 minutes, this amounts to an hourly wage of £9.48).

In line with our preregistered exclusion criteria, we excluded four participants who stated that their data should not be used, and two participants who experienced technical issues. There were no participants who showed clear signs of inattentive responding (e.g., all participants took longer for their participation than our preregistered exclusion criterion of 15 minutes).

\subsubsection{Study design}

\begin{figure}[t]
\begin{subfigure}{0.45\linewidth}
\includegraphics[width=\linewidth]{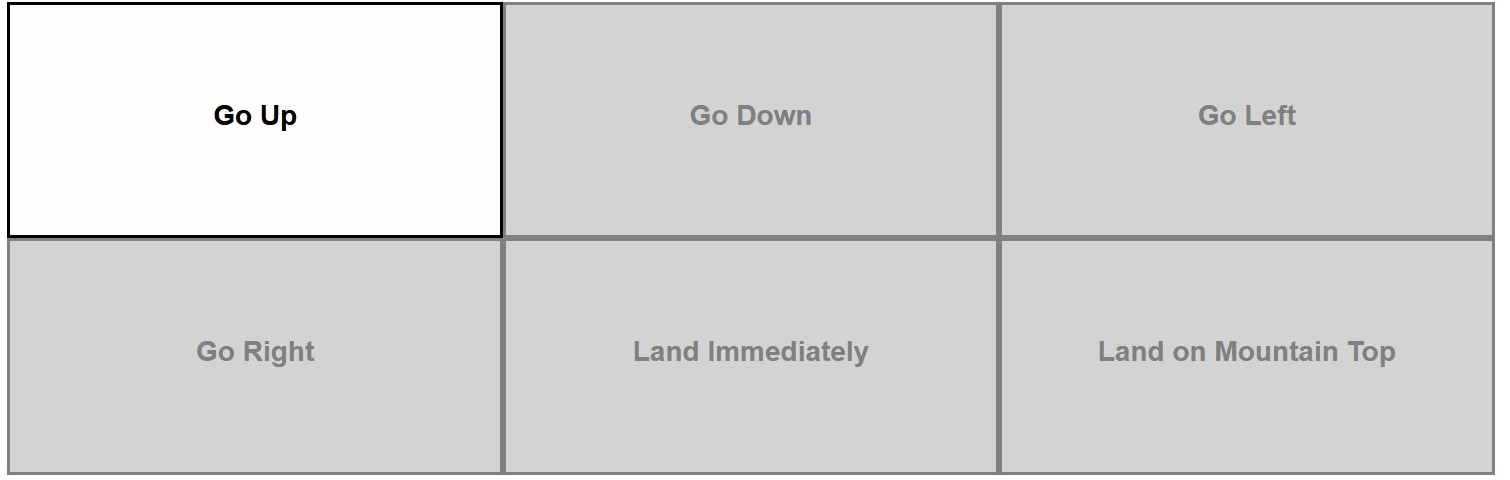}
\caption{\emph{One} \textit{Selectable Actions}}
\label{fig:1Condition}
\end{subfigure}
\begin{subfigure}{0.45\linewidth}
\includegraphics[width=\linewidth]{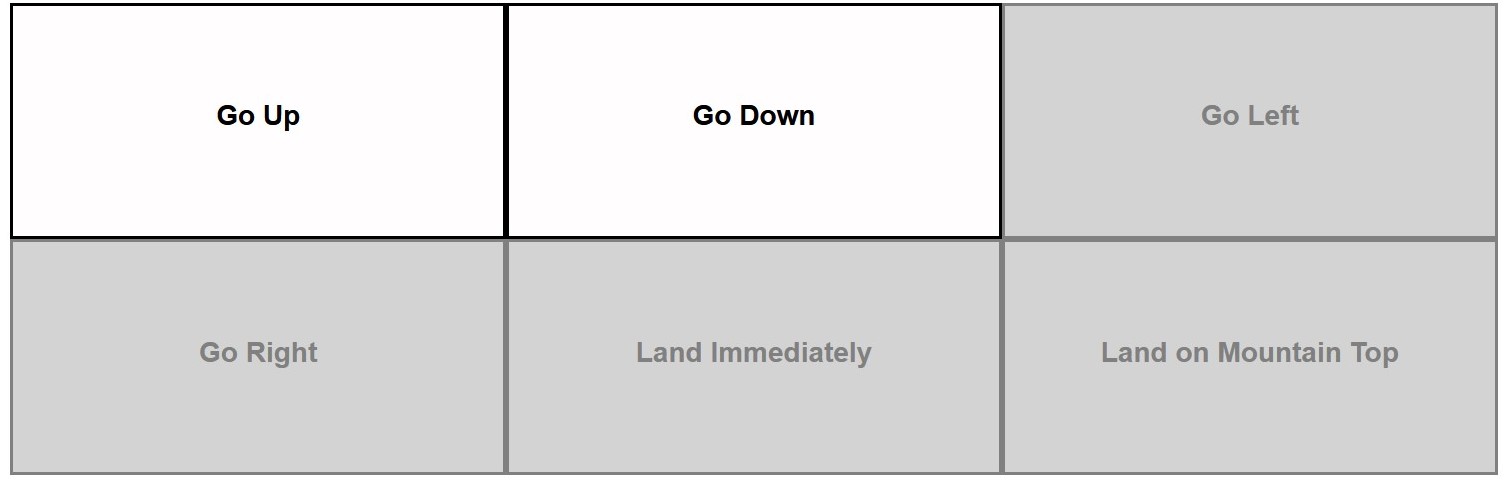}
\caption{\emph{Two} \textit{Selectable Actions}}
\label{fig:2Condition}
\end{subfigure}
\begin{subfigure}{0.45\linewidth}
\includegraphics[width=\linewidth]{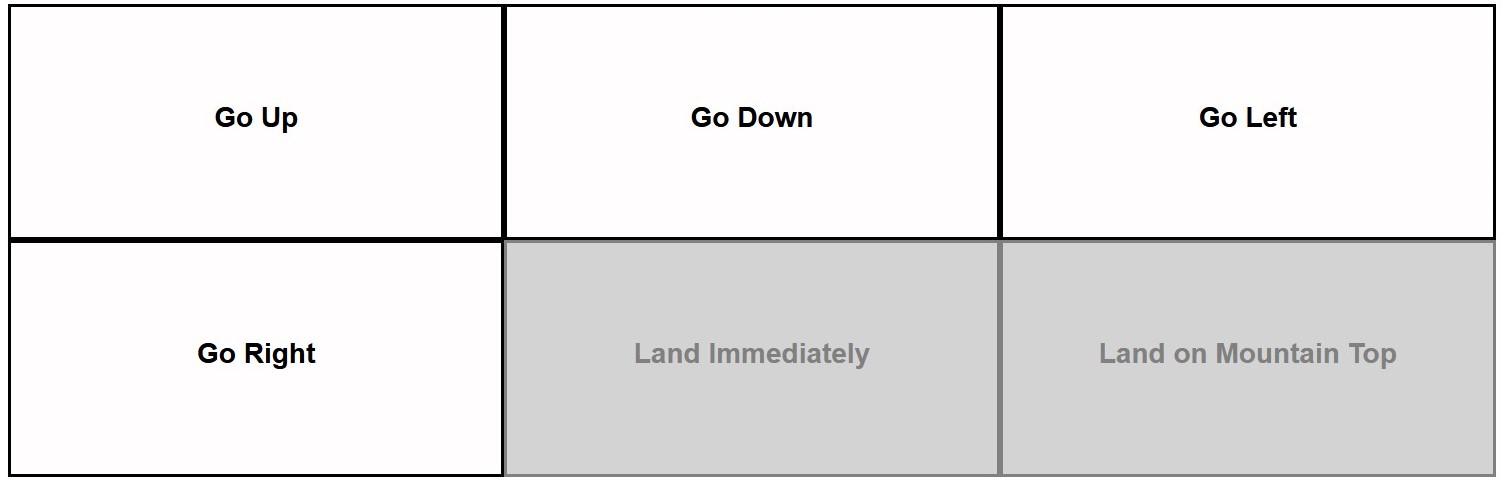}
\caption{\emph{Four} \textit{Selectable Actions}}
\label{fig:4Condition}
\end{subfigure}
\begin{subfigure}{0.45\linewidth}
\includegraphics[width=\linewidth]{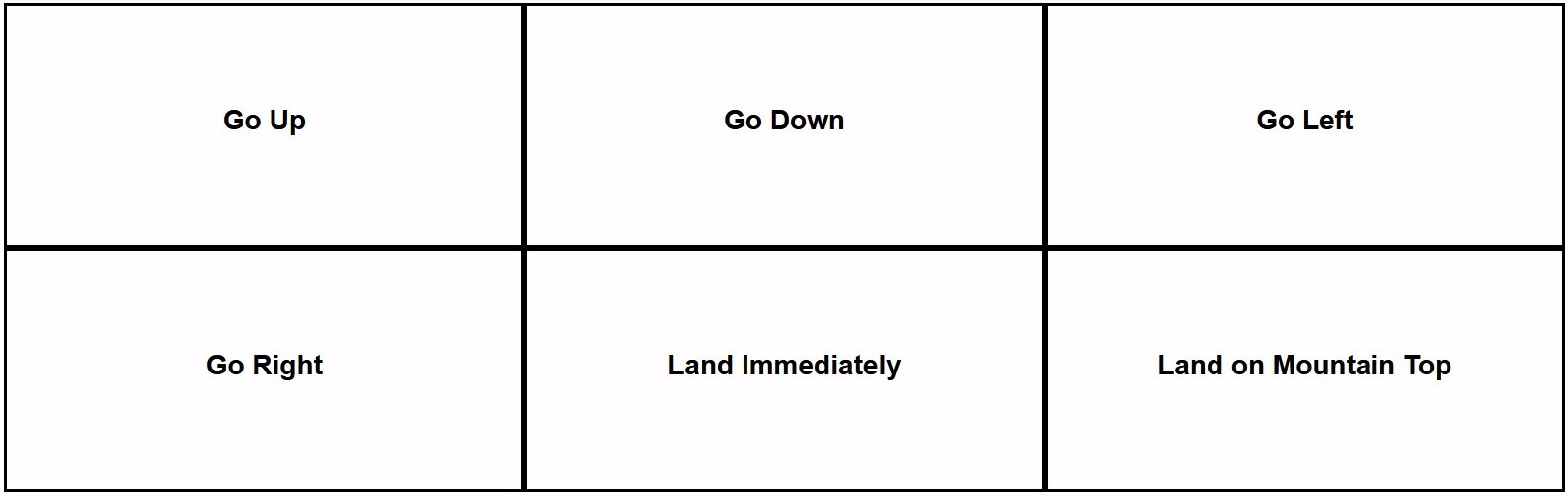}
\caption{\emph{Six} \textit{Selectable Actions}}
\label{fig:6Condition}
\end{subfigure}
\caption{Differences in the decision-making user interface depending on experimental conditions.}
\label{fig:conditions}
\Description{Differences of the decision-making User Interface depending on the four experimental conditions: a) one selectable action, b) two selectable actions, c) four selectable actions, and d) six selectable actions. Each of these User Interfaces consisted of two rows with three rectangular buttons each. The buttons are labeled from top left to bottom right: "Go Up", "Go Down", "Go Left", "Go Right", "Land Immediately", and "Land on Mountain Top". In subfigure a) the button labeled "Go Up" is white with black text, and the other buttons are gray with darker gray text. In subfigure b) the button labeled "Go Up" and "Go Down" are white with black text, and the other buttons are gray with darker gray text. In subfigure c) the button labeled "Go Up", "Go Down", "Go Left", and "Go Right" are white with black text, and the other buttons are gray with darker gray text. In subfigure d) all buttons are white with black text.}
\end{figure}

The experiment was a between-participants study with \textit{Selectable Actions} as the experimental factor. The different levels of \textit{Selectable Actions} are \emph{One}, \emph{Two}, \emph{Four}, and \emph{Six} \textit{Selectable Actions}. For each handover scenario, the participants saw six possible actions, but depending on the experimental condition some of them were grayed out and not selectable, as shown in \autoref{fig:conditions}. The action that successfully resolved the situation was always available. Accordingly, in condition \emph{One}, the only \textit{Selectable Action} would always resolve the critical situation successfully. 
Actions selectable in conditions \emph{Four} and \emph{Two} were selected based on outcomes from the study performed by \citet{gundappa2024designinformativetakeoverrequests}, where the other \textit{Selectable Actions} were the ones most often chosen by participants. 

\subsection{Task and Procedure} 
Participants watched ten 40s-videos of simulated drones flying autonomously. Their task was to monitor the camera output and displayed sensor values to make an informed choice at the end of each video, when participants had to take control of the drone during a critical situation where they then had to choose a suitable action that would avoid a crash. 

\begin{figure}[t]
    \centering
    \includegraphics[width=1\linewidth]{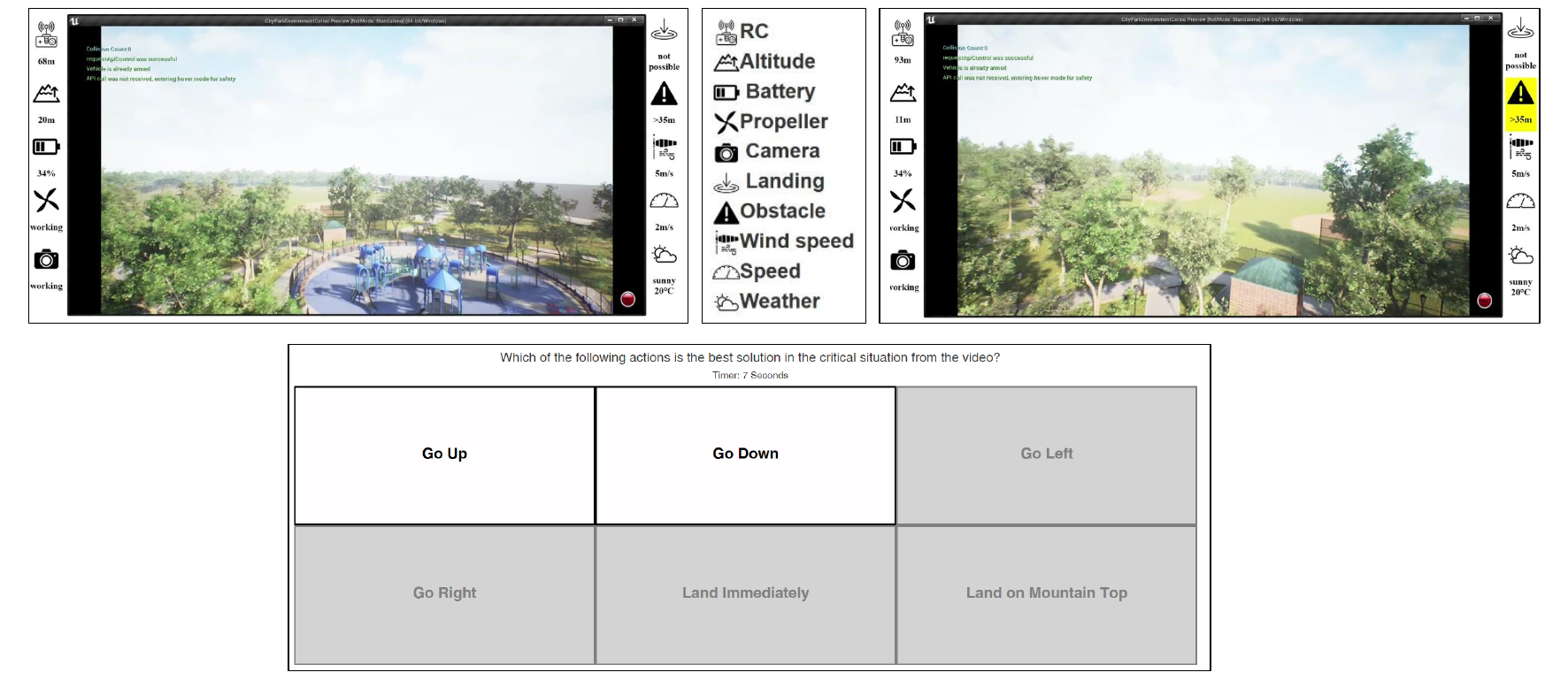}
    \caption{Screenshots of the Drone Monitoring Interface. On the top left, a screenshot of the demo task is shown, where participants see a video recorded by the drone and information about the current status of the drone. In the top middle, the legend of all icons which were used in the interface is shown. This legend was shown to the participants during the entire video of the drone. Ten seconds before the participants had to make a decision, the drone entered a critical situation, which was indicated by an auditory signal and highlighted icons for critical values, as shown in the top right. After these ten seconds, the participants saw six possible actions, and depending on the experimental condition some of them were grayed out indicating they were not available as seen in the bottom screenshot of the \emph{Two} \textit{Selectable Actions} condition. In this demo task, the critical situation was the drone approaching trees which could be resolved by going up.}
    \label{fig: task}
    \Description{
    Screenshots of the Drone Monitoring Interface. 
    In the top middle, there is a legend labeling each of the icons used in the interface: RC, Altitude, Battery, Propeller, Camera, Landing, Obstacle, Wind Speed, Speed, and Weather. This legend was shown to the participants during the entire video of the drone. 
    On the top left, a screenshot of the demo task is shown, where participants see a video recorded by the drone and information about the current status of the drone. The screenshot shows a video from the view of a drone flying above a playground in a virtual environment. To the left and right of this video, each of the icons from the legend is shown with some example values. 
    On the top left the same interface is shown as in the top left, but now the drone approaches some trees. Some of the values changed and the distance icon is highlighted with a yellow background. This screenshot shows a critical situation, which the drone enters in the last ten seconds of each task. In the bottom the decision-making interface is shown. On the top it asks "Which of the following actions is the best solution in the critical situation from the video?" and shows a timer counting down from seven. The main part of the interface are two rows, consisting of three buttons each. The buttons are labeled from top left to bottom right: "Go Up", "Go Down", Go Left", "Go Right", "Land Immediately", and "Lad on Mountain Top". The buttons "Go Up" and "Go Down" are white with black text, while the others are gray with darker gray text indicating that they are not selectable.
    }
\end{figure}

\autoref{fig: task} shows the monitoring interface and video feed of the drone. Sensor values included: the distance to the remote control, the current altitude and battery level, if the propeller and camera are working, if it is possible to land immediately, the distance to the next obstacle, the current wind speed and speed of the drone, and the current weather condition.

Toward the end of each video, the drone was in a critical situation that required a handover to the human overseer, which was indicated by an auditory signal. At this point, the video highlighted the sensor values related to the critical situation. After ten seconds a second auditory signal was played, the video disappeared and six actions were displayed (see \autoref{fig: task}, bottom): fly higher, fly lower, fly to the left, fly to the right,  land immediately, or land at a (varying) location. Depending on the condition some of the actions were grayed out (see below). 
Participants were instructed to choose an action as quickly as possible and had up to seven seconds to decide.
To resolve the critical situation successfully, participants had to consider the corresponding sensor values and video feed to consider the possibility of landing at specific locations.

There were three types of critical situations: weather conditions, technical problems, and physical obstacles. Weather conditions caused a critical situation if the wind speed passed a certain threshold (unknown to participants) or if the current weather was rainy, snowy, windy, or foggy and the altitude was too high or too low. Technical problems occurred when the camera or propeller stopped working or the battery was too low. Physical obstacles led to critical situations when the distance to the next obstacle was too little or flying at the current altitude was dangerous.  

After deciding on an action, participants were asked to fill out three open-ended questions to justify their answer (see supp. material). These ensured that people did not choose actions at random. If they did not choose an action, they were asked to give a reason. After that they were provided with feedback stating if their chosen action saved the drone or why it crashed. 

After the 1st, 5th, 9th, and 10th (last) trial, participants filled out a questionnaire with Likert-scale items assessing their perceived autonomy, meaningfulness, task motivation, work satisfaction, and mental demand +  temporal demand (which we combined to users' task load).
In the final trial, there was no action that resolved the critical situation, always leading to a crash. This was to ensure that participants experienced at least one crash so that we could additionally ask questions related to their perceived responsibility in the last questionnaire. Table~\ref{tab:questionnaire_items} shows the items in the questionnaire relevant to this paper. See \autoref{tab:additional_questionnaire_items} in the Appendix for the full set of questions asked, including additional exploratory variables, such as stress, trust, and responsibility-related variables. Since our analysis focused on the perceived autonomy and its impact on motivational variables, we restricted our results to the variables most relevant for these goals.

In the beginning of the experiment, participants received written instructions about the task, the meaning of sensor values, the types of critical situations, and possible actions. After the participants had all the required information, they were instructed to solve a trial task. In the trial task, they had the opportunity to make their own decision. Regardless of their decision, they got to know which action would have resolved the critical situation successfully. We also provided them with examples for this task on how to fill the open-ended questions.  Exact instructions and screenshots from all steps of the experiment are provided in the supplementary material. 


\subsection{Implementation and Video material}
The study we created was based on a previous study performed by \citet{gundappa2024designinformativetakeoverrequests}. They used their simulation framework that extends Microsoft AirSim~\cite{shah2018airsim} and can be used to create a variety of handover situations. In total, they created 20 different scenarios using two existing AirSim environments (i.e, park and valley environment). In our study, we used ten scenarios that had only one action that resolved the critical situation successfully. We made this choice so that there is room for wrong decisions in the \emph{Two} and \emph{Four} Conditions, while we did not want to make the participants lose trust in the recommendations by making correct actions not selectable. One of the scenarios is provided in the supplementary materials. We embedded the videos in a web page, which enabled us to create this interactive task and recruit participants on the Prolific Platform.


\subsection{Measured Variables}

We used self-report items to measure perceived \textit{autonomy}, \textit{meaningfulness}, \textit{task motivation}, \textit{task satisfaction} and \textit{task load} on a 7-point Likert scale (1 - strongly disagree to 7 - strongly agree). The items we used are shown in \autoref{tab:questionnaire_items} along with references to the literature from which we sourced these items. All additionally measured variables not included in this paper are shown in \autoref{tab:additional_questionnaire_items} in the Appendix. 
Self-report items were measured multiple times (after the 1., 5., 9., final scenario). 

Note that given the substantial overall correlation between task motivation and task satisfaction (r = .87), for the remainder of this paper, we decided to only report task motivation because the findings for task satisfaction were nearly identical.

Additionally, for all ten scenarios, we measured two decision-related variables: \textit{decision accuracy} and \textit{decision time}. Decision accuracy was calculated for each participant as the percentage of selected actions that led to a successful outcome (i.e. drone not crashing) in the first nine rounds. If a participant did not select an action, this was counted as inaccurate. The last scenario was excluded because all actions led to a crash. 
Decision time was measured as the time after participants saw the available actions until they chose a specific action.

\begin{table*}[t]
    \centering
        \caption{An overview of the variables and corresponding questionnaire items collected and analyzed in this paper, the rounds after which each question was asked, and the sources from which they were adapted. }
\label{tab:questionnaire_items}
    \begin{tabular}{>{\raggedright\arraybackslash}p{0.3\linewidth} >{\raggedright\arraybackslash}p{0.5\linewidth} c c} \toprule
        Concept & Question(s)&  Ref\\ \midrule
        Perceived Autonomy & 'I am so restricted by the available actions that I can hardly make my own decisions.' \textit{(inversely coded)} &  \cite{van_dick_job_2001}\\
        &  'I can make decisions about how to prevent crashes independently.'&\\ 
        Perceived Meaningfulness & 'The work I do is important.'  &\cite{spreitzer_empirical_1995}\\
        & 'The work I do is meaningful.'& \\ 
        Task Motivation&'I have fun performing the task.'  &\cite{gagne_motivation_2010}\\
        Task Satisfaction&'I feel fulfilled performing the task.' &\cite{parker_measurement_2011}\\
        Task Load&'The task was mentally demanding.'    &\cite{hart_development_1988}\\
        &'The pace of the task was hurried or rushed.' & \\
        \bottomrule
    \end{tabular}

\end{table*}

\section{Results}

\subsection{Analysis}
We decided to split our analysis into three parts, each requiring different approaches for statistical analysis. In \autoref{sec:AnalysisImpactConditions}, we report the overall effects of \textit{Selectable Actions} on our dependent variables, averaging results from all scenarios and repeated questionnaires. To test for significant differences, we chose a Kruskal-Wallis test combined with Dunn test with Bonferroni correction for the post-hoc comparisons of all experimental conditions. In \autoref{sec:AnalysisOverTime}, we were interested in understanding how \textit{Selectable Actions} impacted decision-related variables, perceived autonomy, and motivational factors over time. To this end, we treated \emph{Decision Round} as an additional independent variable and report results from linear mixed models that test for interaction effects of the \textit{Selectable Actions} and the \emph{Decision Round}. Finally, in \autoref{sec:AnalysisMotivation}, we combined the data of those conditions where participants had at least two actions to choose from (i.e. \emph{Two}, \emph{Four}, and \emph{Six} (informed by our results in \autoref{sec:AnalysisImpactConditions}
). 
Our goal was to analyze the effects of perceived autonomy, meaningfulness, and task motivation on decision accuracy and decision time. To this end, we report results from linear mixed models where perceived autonomy, meaningfulness, and motivation were added as predictors beyond \textit{Selectable Actions} and the \emph{Decision Round}.

\subsection{The impact of the number of \textit{Selectable Actions}}
\label{sec:AnalysisImpactConditions}

\begin{figure}[t]
\begin{subfigure}{0.45\linewidth}
\includegraphics[width=\linewidth]{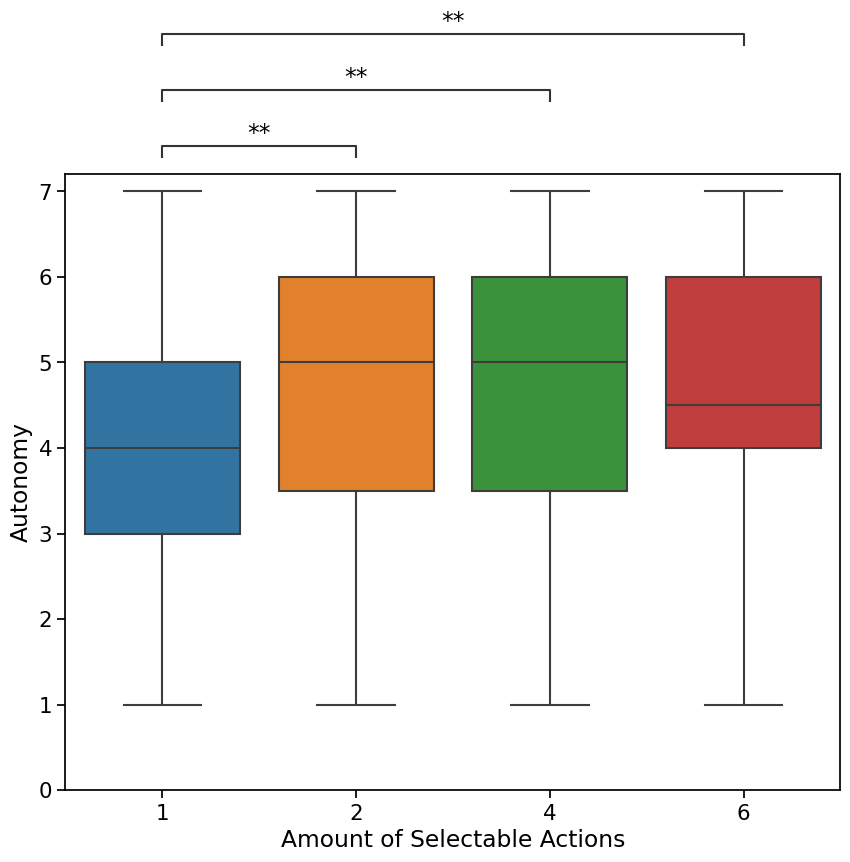}
\caption{Autonomy}
\label{fig:boxplotAutonomy}
\end{subfigure}
\begin{subfigure}{0.45\linewidth}
\includegraphics[width=\linewidth]{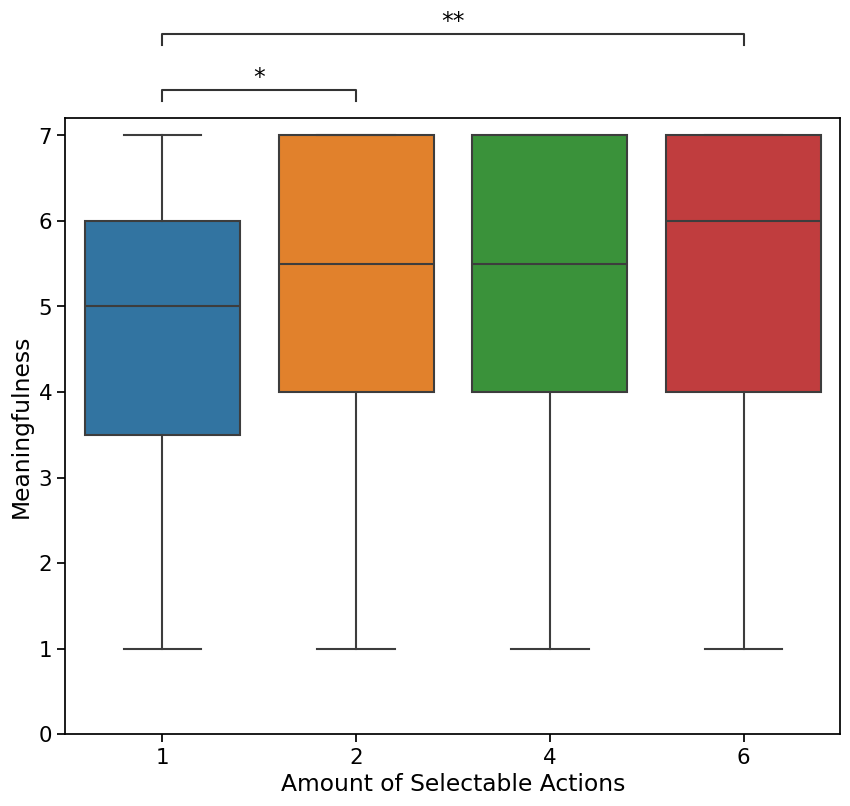}
\caption{Meaningfulness}
\label{fig:boxplotMeaningfulness}
\end{subfigure}
\begin{subfigure}{0.45\linewidth}
\includegraphics[width=\linewidth]{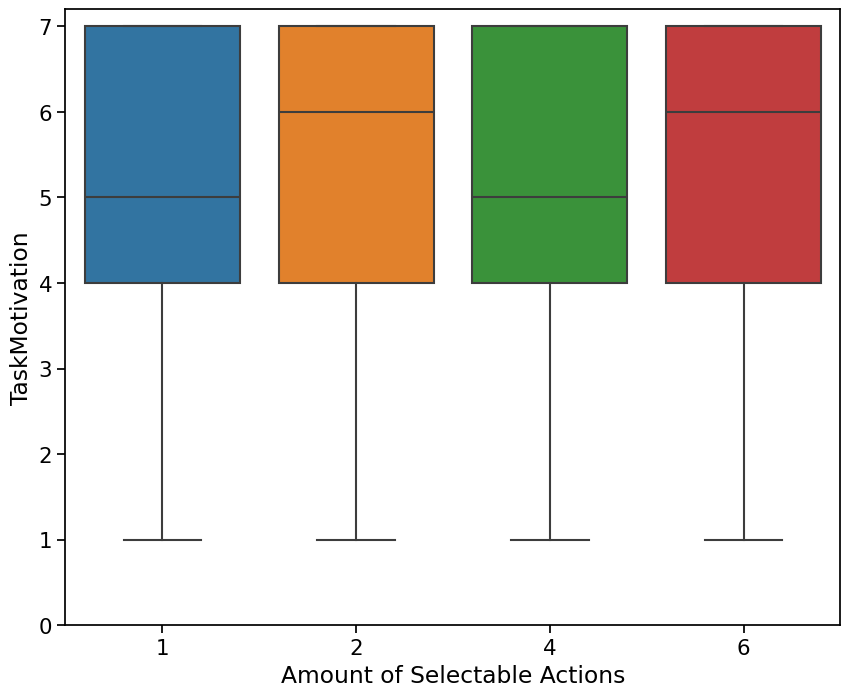}
\caption{Task Motivation}
\label{fig:boxplot_Motivation_Conditions}
\end{subfigure}
\begin{subfigure}{0.45\linewidth}
\includegraphics[width=\linewidth]{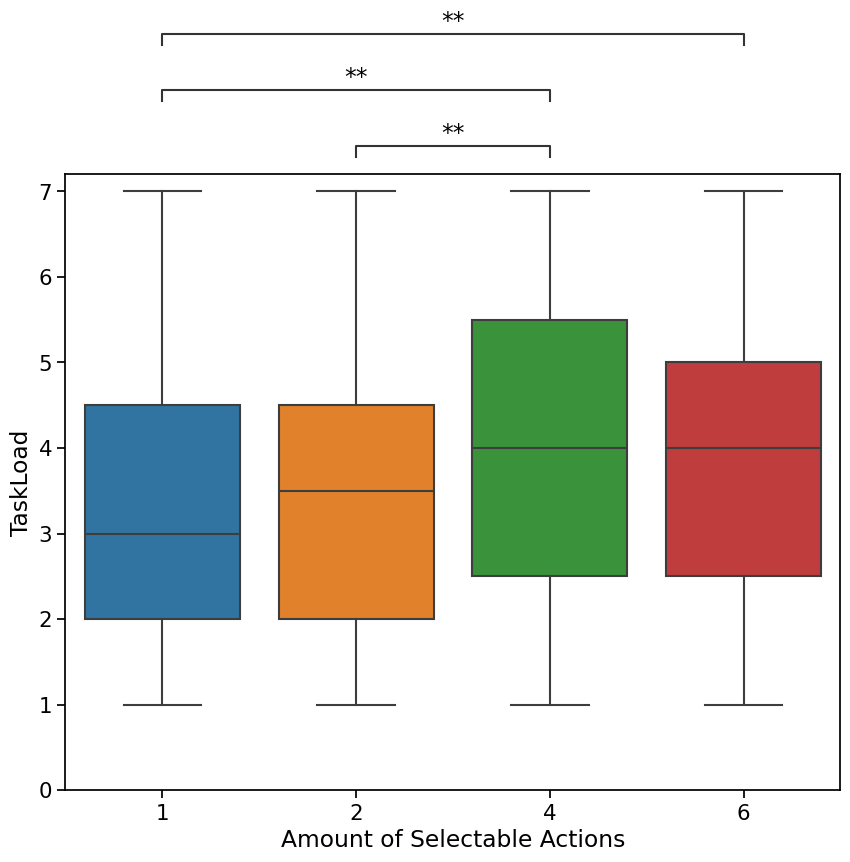}
\caption{Task Load}
\label{fig:boxplotTaskload}
\end{subfigure}\\
Note: *p<=0.05, **p<=0.01
\caption{Differences in perceived autonomy, meaningfulness, task motivation, and task load on a 7-point Likert Scale between the experimental conditions} 
\label{fig:boxplotAutonomyResponsibilityMeaningfulnessTaskload}
\Description{Boxplots of a) Autonomy, b) Meaningfulness, c) Task Motivation, and d) Task Load for the four experimental conditions. 
a) Median values of Autonomy are 4 for one, 5 for two, 5 for four, and 4.5 for six selectable actions. 
Interquartile ranges are 2-2.5.
b) Median values of Meaningfulness are 5 for one, 5.5 for two, 5.5 for four, and 6 for six selectable actions. 
Interquartile ranges are 2.5-3.
c) Median values of task motivation are 5 for one, 6 for two, 5 for four, and 6 for six selectable actions. 
Interquartile ranges are 3 points on a 7-point Likert Scale.
d) Median values of Task Load are 3 for one, 3.5 for two, 4 for four, and 4 for six selectable actions. 
Interquartile ranges are 2.5-3.
}
\end{figure}

\begin{figure}[t]
\begin{subfigure}{0.45\linewidth}
\includegraphics[width=\linewidth]{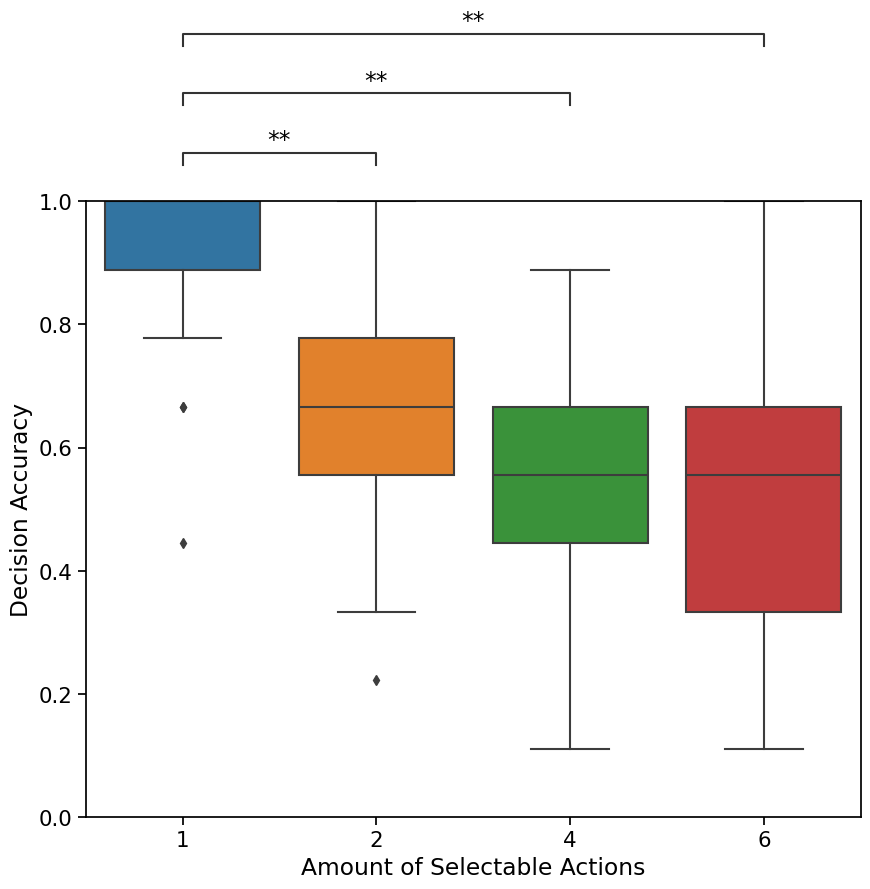}
\caption{Decision Accuracy}
\label{fig:boxplot_DecisionAccuracy_Conditions}
\end{subfigure}
\begin{subfigure}{0.45\linewidth}
\includegraphics[width=\linewidth]{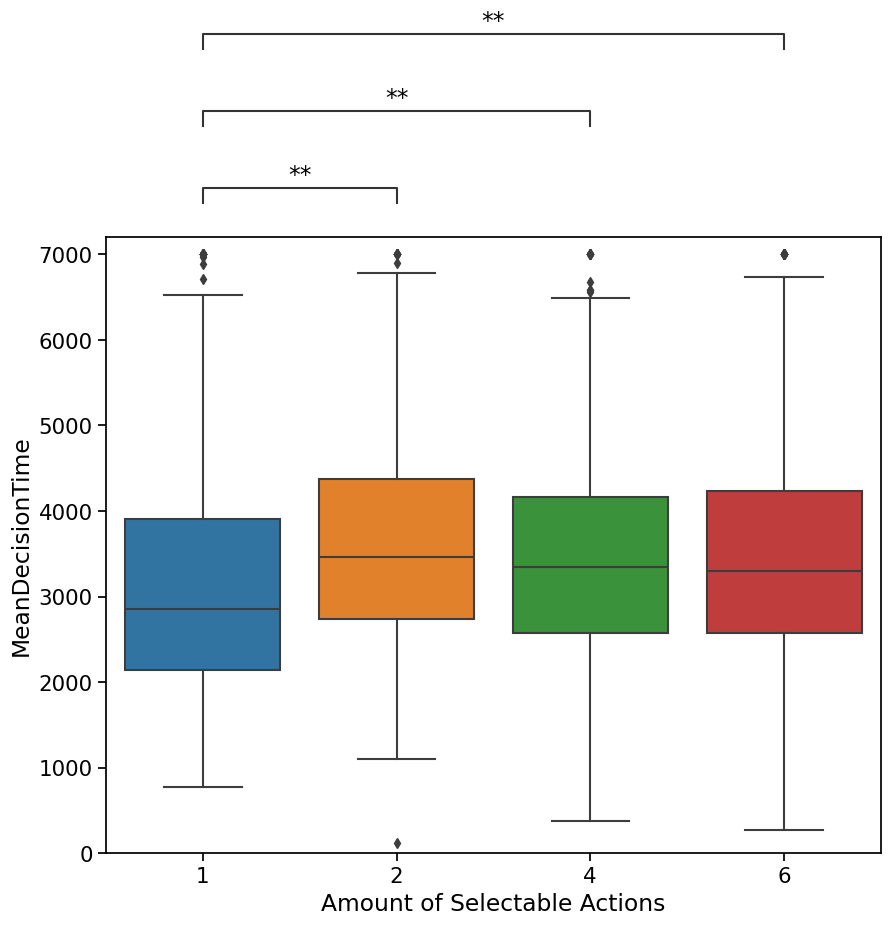}
\caption{Decision Time in ms}
\label{fig:boxplot_DecisionTime_Conditions}
\end{subfigure}\\
Note: *p<=0.05, **p<=0.01 
\caption{Differences in Decision Accuracy in Percentage and Decision Time in milliseconds between the Experimental Conditions}
\label{fig:boxplotDecisionTimeAccuracy}
\Description{Boxplots of a) Decision Accuracy and b) Decision Time for the four experimental conditions. 
a) Median values of Decision Accuracy are 1 for one, 0.66 for two, 0.55 for four, and 0.55 for six selectable actions. 
Interquartile ranges are 0.11-0.33.
b) Median values of Meaningfulness are 2851 ms for one, 3268 ms for two, 3386 ms for four, and 3289 ms for six selectable actions. 
Interquartile ranges are 721ms-1079ms.
}
\end{figure}

Research Questions 1a-f asked about the effects of \emph{Selectable Actions} on various decision-related and self-report variables. 

\emph{Selectable Actions} significantly affected participants' perceived autonomy (H~=~38.35, p~<~.001), meaningfulness (H~=~15.52, p~=~.001), and task load (H~=~30.71, p~<~.001). 
As visualized by the box-plots in \autoref{fig:boxplotAutonomyResponsibilityMeaningfulnessTaskload}, with \emph{One} \textit{Selectable Action}, participants felt less autonomous than having \emph{Two} (p~<~.001), \emph{Four} (p~<~.001), or \emph{Six} (p~<~.001).  
Participants judged the task as less meaningful when having \emph{One} \textit{Selectable Action} compared to having \emph{Two} (p~=~.02), or \emph{Six} (p~=~.001).  Also, with \emph{One} \textit{Selectable Action} they had a lower task load than with \emph{Four} (p~<~.001), or \emph{Six} (p~<~.001). In contrast, there was no significant difference for  Motivation across all conditions (see \autoref{fig:boxplot_Motivation_Conditions}).

The number of \emph{Selectable Actions} also significantly affected the decision accuracy (\autoref{fig:boxplot_DecisionAccuracy_Conditions}) (H~=~128.88, p~<~.001). Pairwise post-hoc comparisons showed participants having \emph{One} \textit{Selectable Action} to make better decisions than participants having \emph{Two} (p~<~.001), \emph{Four} (p~<~.001), or \emph{Six} (p~<~.001) (\autoref{fig:boxplot_DecisionAccuracy_Conditions}). Although visually there is a trend that decision accuracy decreased with an increasing number of selectable actions, no significant differences were found when comparing conditions with more than one action to choose from.

We also found a significant difference in the \textit{decision time} between the experimental conditions (H~=~28.79, p~<~.001). Participants who had \emph{One} \textit{Selectable Action} decided faster than those with \emph{Two} (p~<~.001), \emph{Four} (p~<~.001), or \emph{Six} (p~<~.001) (\autoref{fig:boxplot_DecisionTime_Conditions}). No significant differences were found when comparing conditions with more than one action to choose from.

\subsection{Changes in the effects of \textit{Selectable Actions} over time}
\label{sec:AnalysisOverTime}
The findings up to now indicate that there are general differences between the experimental conditions but mostly between the condition with \emph{One} \textit{Selectable Action} and the conditions that had at least some available choice for an action. Given that the effects of varying \textit{Selectable Actions} may only unfold over time or may grow stronger over time, we decided to additionally examine interactive effects of the \emph{Selectable Actions} and the \emph{Decision Round}, that is after how many scenarios the questionnaire was asked (1st, 5th, 9th, or 10th). 

\begin{figure}[t]
\begin{subfigure}{0.24\linewidth}
\includegraphics[width=\linewidth]{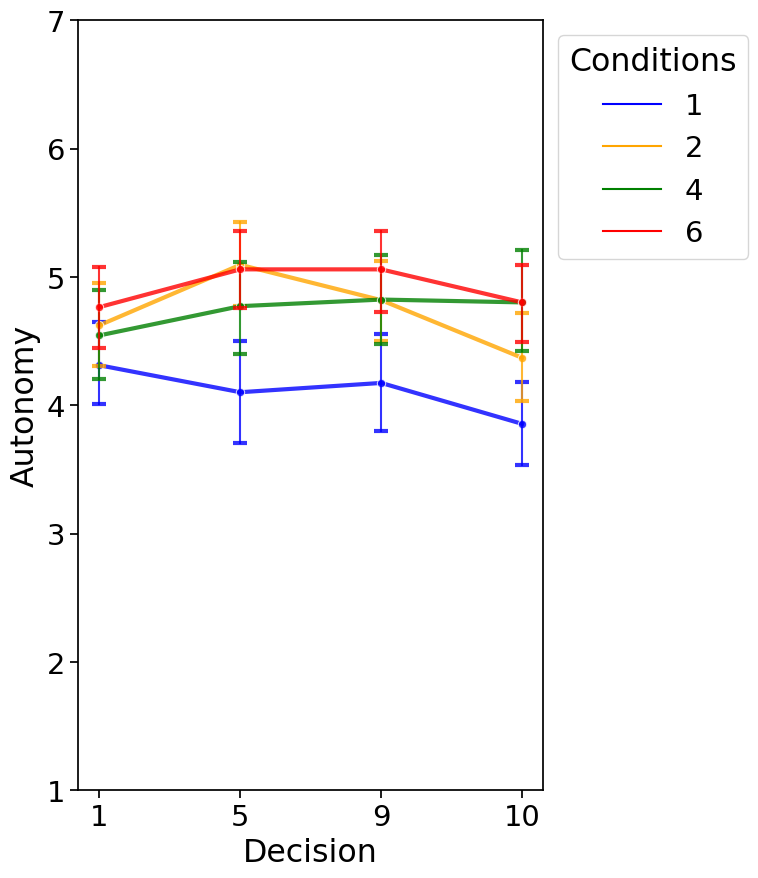}
\caption{Autonomy}
\label{fig:boxplotAutonomyPointOfTime}
\end{subfigure}
\begin{subfigure}{0.24\linewidth}
\includegraphics[width=\linewidth]{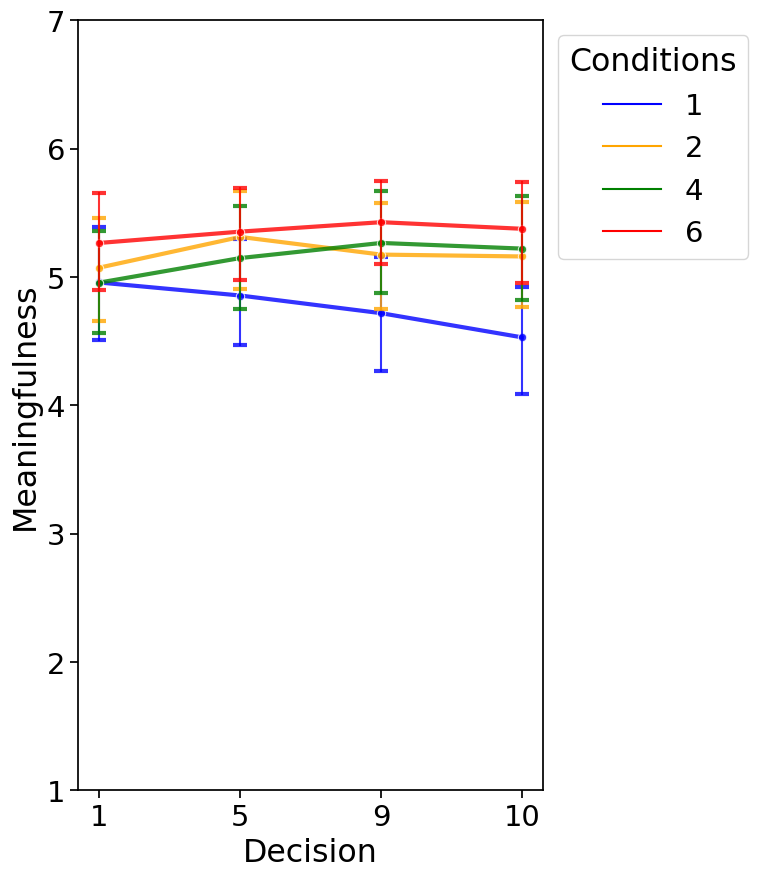}
\caption{Meaningfulness}
\label{fig:boxplotMeaningfulnessPointOfTime}
\end{subfigure}
\begin{subfigure}{0.24\linewidth}
\includegraphics[width=\linewidth]{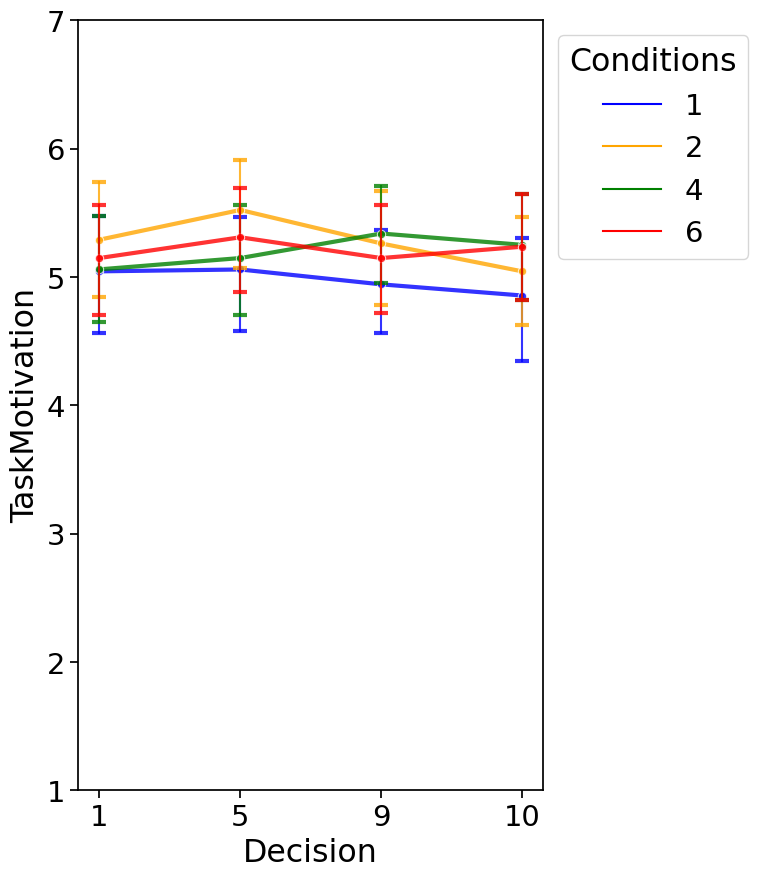}
\caption{Motivation}
\label{fig:boxplotMotivationPointOfTime}
\end{subfigure}
\begin{subfigure}{0.24\linewidth}
\includegraphics[width=\linewidth]{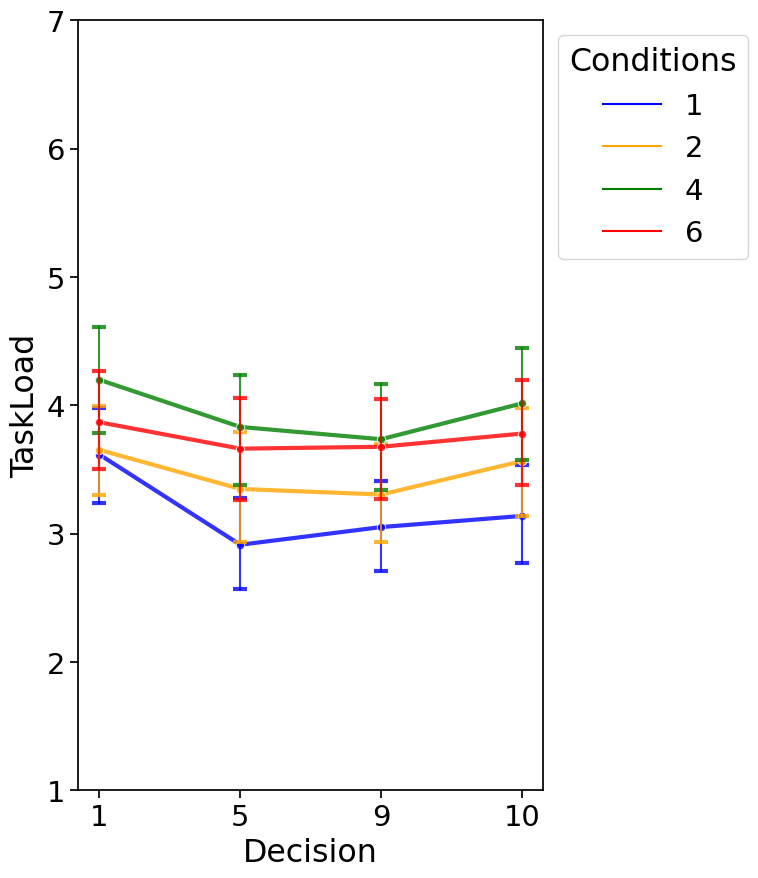}
\caption{Task Load}
\label{fig:boxplotTaskloadPointOfTime}
\end{subfigure}
\Description{Lineplots of a) Autonomy, b) Meaningfulness, c) Task Motivation, and d) Task Load for the four points of time and experimental conditions. 
a) 
Mean values of Autonomy after Decision 1 are 4.3 for one, 4.6 for two, 4.5 for four, and 4.7 for six selectable actions. 
The size of the 0.95 confidence intervals is 0.6-0.8.
Mean values of Autonomy after Decision 5 are 4.1 for one, 5 for two, 4.7 for four, and 5 for six selectable actions. 
The size of the 0.95 confidence intervals is 0.6-0.9.
Mean values of Autonomy after Decision 9 are 4.1 for one, 4.8 for two, 4.8 for four, and 5 for six selectable actions. 
The size of the 0.95 confidence intervals is 0.6-0.8.
Mean values of Autonomy after Decision 10 are 3.8 for one, 4.3 for two, 4.8 for four, and 4.8 for six selectable actions. 
The size of the 0.95 confidence intervals is 0.6-0.7.
b) 
Mean values of Meaningfulness after Decision 1 are 4.9 for one, 5 for two, 4.9 for four, and 5.2 for six selectable actions. 
The size of the 0.95 confidence intervals is 0.8-1.
Mean values of Meaningfulness after Decision 5 are 4.8 for one, 5.3 for two, 5.1 for four, and 5.3 for six selectable actions. 
The size of the 0.95 confidence intervals is 0.7-0.9.
Mean values of Meaningfulness after Decision 9 are 4.7 for one, 5.1 for two, 5.2 for four, and 5.4 for six selectable actions. 
The size of the 0.95 confidence intervals is 0.7-0.9.
Mean values of Meaningfulness after Decision 10 are 4.5 for one, 5.1 for two, 5.2 for four, and 5.3 for six selectable actions. 
The size of the 0.95 confidence intervals is 0.7-1.
c)
Mean values of Task Motivation after Decision 1 are 5 for one, 5.2 for two, 5 for four, and 5.1 for six selectable actions. 
The size of the 0.95 confidence intervals is 0.8-1.
Mean values of Task Motivation after Decision 5 are 5 for one, 5.5 for two, 5.1 for four, and 5.3 for six selectable actions. 
The size of the 0.95 confidence intervals is 0.8-1.
Mean values of Task Motivation after Decision 9 are 4.9 for one, 5.2 for two, 5.3 for four, and 5.1 for six selectable actions. 
The size of the 0.95 confidence intervals is 0.8-0.9.
Mean values of Task Motivation after Decision 10 are 4.8 for one, 5 for two, 5.2 for four, and 5.2 for six selectable actions. 
The size of the 0.95 confidence intervals is 0.8-1.
d)
Mean values of Task Load after Decision 1 are 3.6 for one, 3.6 for two, 4.1 for four, and 3.8 for six selectable actions. 
The size of the 0.95 confidence intervals is 0.7-0.8.
Mean values of Task Load after Decision 5 are 2.9 for one, 3.3 for two, 3.8 for four, and 3.6 for six selectable actions. 
The size of the 0.95 confidence intervals is 0.7-0.9.
Mean values of Task Load after Decision 9 are 3 for one, 3.3 for two, 3.7 for four, and 3.6 for six selectable actions. 
The size of the 0.95 confidence intervals is 0.7-0.9.
Mean values of Task Load after Decision 10 are 3.1 for one, 3.5 for two, 4 for four, and 3.7 for six selectable actions. 
The size of the 0.95 confidence intervals is 0.8-0.9.
}
    
\caption{Differences in perceived autonomy, meaningfulness, task motivation, and task load over time in each condition on a 7-point Likert Scale} 
\label{fig:boxplotPointOfTime}
\end{figure}

\autoref{fig:boxplotPointOfTime} and \autoref{tab:hlmTime} show the results of linear mixed model analyses with the dependent variables perceived autonomy, meaningfulness, motivation, task load, and decision time. The decision round is nested within participants as every participant experienced 10 decision rounds where we captured the relevant dependent variables four times. As predictors in the linear mixed models we included a) the \emph{Selectable Actions} as a dummy-coded variable with \emph{One} \textit{Selectable Action} as the reference condition, b) the four \emph{Decision Rounds} with Decision 1 as the reference condition, and c) the interaction effect between the \emph{Selectable Actions} and the \emph{Decision Round}. 


For perceived autonomy, \autoref{fig:boxplotAutonomyPointOfTime} together with the significant interaction effects between the \emph{Decision Round} and the \textit{Selectable Actions} reported in \autoref{tab:hlmTime} show that over time, the differences in perceived autonomy become larger. While at decision 1, there were comparably small differences for \emph{Selectable Actions}, the differences between \emph{One} \textit{Selectable Action} and the other conditions were larger for later rounds.

For perceived meaningfulness, only for the ninth and for the final round were there significant interaction effects (see \autoref{tab:hlmTime}). This indicates that over time the difference in perceived meaningfulness between the condition with \emph{One} \textit{Selectable Action} and the other conditions became larger (see \autoref{fig:boxplotMeaningfulnessPointOfTime}).

For task motivation (\autoref{fig:boxplotMotivationPointOfTime}) and task load (\autoref{fig:boxplotTaskloadPointOfTime}) there were no clear effects other than a main effect for the \emph{Decision Round} indicating a lower perceived task load for later rounds.

For the decision time, there were also no clear effects other than participants taking less time for the later rounds than for the first round.

\begin{table}[t] 
\small
\centering 
  \caption{Linear mixed model analyses with the dependent variables perceived
autonomy, meaningfulness, motivation, task load, and decision time} 
  \label{tab:hlmTime} 
\begin{tabular}{@{\extracolsep{5pt}}lccccc} 
\hline \\[-1.8ex] 
 & \multicolumn{5}{c}{\textit{Dependent variable:}} \\ 
\cline{2-6} 
\\[-1.8ex] & Autonomy & Meaningfulness & TaskMotivation & TaskLoad & MeanDecisionTime \\ 
\hline \\[-1.8ex] 
 Intercept & \textbf{4.31$^{**}$ (0.18)} & \textbf{4.96$^{**}$ (0.21)} & \textbf{5.04$^{**}$ (0.22)} & \textbf{3.29$^{**}$ (0.15)} & \textbf{4,011.33$^{**}$ (163.79)} \\ 
  2 Actions & 0.31 (0.25) & 0.12 (0.29) & 0.25 (0.31) & 0.09 (0.21) & \textbf{473.12$^{*}$ (231.63)} \\ 
  4 Actions & 0.23 (0.25) & $-$0.00 (0.30) & 0.02 (0.31) & 0.29 (0.21) & $-$97.10 (232.48) \\ 
  6 Actions & 0.45 (0.25) & 0.31 (0.30) & 0.10 (0.31) & 0.28 (0.21) & 56.87 (232.48) \\ 
  Decision 5 & $-$0.21 (0.15) & $-$0.10 (0.13) & 0.01 (0.17) & \textbf{$-$0.23$^{**}$ (0.05)} & \textbf{$-$1,101.45$^{**}$ (210.27)} \\ 
  Decision 9 & $-$0.14 (0.15) & $-$0.24 (0.13) & $-$0.10 (0.17) & \textbf{$-$0.19$^{**}$ (0.05)} & \textbf{$-$1,141.52$^{**}$ (210.27)} \\ 
  Decision 10 & \textbf{$-$0.46$^{**}$ (0.15)} & \textbf{$-$0.43$^{**}$ (0.13)} & $-$0.19 (0.17) & \textbf{$-$0.16$^{**}$ (0.05)} & \textbf{$-$931.03$^{**}$ (210.27)} \\ 
  2 Actions:Decision 5 & \textbf{0.68$^{**}$ (0.21)} & 0.34 (0.18) & 0.22 (0.24) & 0.13 (0.07) & $-$157.15 (297.36) \\ 
  4 Actions:Decision 5 & \textbf{0.44$^{*}$} (0.22) & 0.29 (0.18) & 0.07 (0.24) & 0.11 (0.07) & 417.74 (298.45) \\ 
  6 Actions:Decision 5 & \textbf{ 0.50$^{*}$} (0.22) & 0.19 (0.18) & 0.15 (0.24) & \textbf{0.17$^{*}$ (0.07)} & 289.13 (298.45) \\ 
  2 Actions:Decision 9 & 0.33 (0.21) & 0.34 (0.18) & 0.07 (0.24) & 0.07 (0.07) & 287.27 (297.36) \\ 
  4 Actions:Decision 9 & 0.42 (0.22) & \textbf{0.55$^{**}$ (0.18)} & 0.38 (0.24) & 0.03 (0.07) & \textbf{706.60$^{*}$ (298.45)} \\ 
  6 Actions:Decision 9 & \textbf{0.43$^{*}$ (0.22)} & \textbf{0.40$^{*}$ (0.18)} & 0.10 (0.24) & 0.12 (0.07) & \textbf{632.72$^{*}$ (298.45)} \\ 
  2 Actions:Decision 10 & 0.20 (0.21) & \textbf{0.51$^{**}$ (0.18)} & $-$0.06 (0.24) & 0.13 (0.07) & $-$46.01 (297.36) \\ 
  4 Actions:Decision 10 & \textbf{0.71$^{**}$ (0.22)} & \textbf{0.69$^{**}$ (0.18)} & 0.38 (0.24) & 0.10 (0.07) & 447.12 (298.45) \\ 
  6 Actions:Decision 10 & \textbf{0.49$^{*}$ (0.22)} & \textbf{0.54$^{**}$ (0.18)} & 0.28 (0.24) & 0.13 (0.07) & 398.21 (298.45) \\ 
 \hline \\[-1.8ex] 
Observations & 1,096 & 1,096 & 1,096 & 1,096 & 1,096 \\ 
\hline \\[-1.8ex] 
\multicolumn{6}{l} {\textit{Note:} The reference condition for Actions is 1 Action; the reference condition for Decision is Decision 1. $^{*}$p$<$0.05; $^{**}$p$<$0.01}\\ 
\end{tabular} 
\end{table}

Building on the results up to this point, we can respond to the Research Questions 1a-1f:
The number of \emph{Selectable Actions}…
\begin{enumerate}[label={RQ1\alph*.}]
    \item \textbf{affected} users’ perceived autonomy. The differences between \emph{One} \textit{Selectable Action} and the other conditions was stronger for later rounds than for \emph{Decision Round} 1.
    \item \textbf{affected} users’ perceived meaningfulness. The differences between \emph{One} \textit{Selectable Action} and the other conditions was stronger for later rounds than for \emph{Decision Round} 1.
    \item \textbf{did not affect} users’ task motivation.
    \item \textbf{affected} users’ task load. There was no clear significant interaction effects between the number of \emph{Selectable Actions} and the \emph{Decision Round}.
    \item \textbf{affected} the decision accuracy in the drone handover task. Participants who had \emph{One} \textit{Selectable Action} performed better than participants in the other conditions.
    \item \textbf{affected} users’ decision time. Participants who had \emph{One} \textit{Selectable Action} took less time to decide than participants in the other conditions. There was no clear significant interaction effects between the number of \emph{Selectable Actions} and the \emph{Decision Round}.
\end{enumerate}

\begin{table}[t] 
\centering 
\small
  \caption{Linear mixed models with the dependent variable decision accuracy} 
  \label{tab:hlmDecisionAccuracy} 
\begin{tabular}{@{\extracolsep{5pt}}lcccc} 
\hline \\[-1.8ex] 
 & \multicolumn{4}{c}{\textit{Dependent variable:}} \\ 
\cline{2-5} 
\\[-1.8ex] & \multicolumn{4}{c}{PercentageCorrectAnswers} \\ 
\\[-1.8ex] & (1) & (2) & (3) & (4)\\ 
\hline \\[-1.8ex] 
 Intercept & \textbf{0.25$^{**}$ (0.06)} & \textbf{0.54$^{**}$ (0.06)} & \textbf{0.48$^{**}$ (0.06)} & \textbf{0.55$^{**}$ (0.07)} \\ 
  Autonomy & \textbf{0.05$^{**}$ (0.01)} &  &  &  \\ 
  Meaningfulness &  & $-$0.01 (0.01) &  &  \\ 
  TaskMotivation &  &  & $-$0.00 (0.01) &  \\ 
  TaskLoad &  &  &  & $-$0.02 (0.01) \\ 
  4 Actions & $-$0.02 (0.06) & $-$0.02 (0.06) & $-$0.02 (0.06) & $-$0.02 (0.06) \\ 
  6 Actions & \textbf{$-$0.16$^{**}$ (0.06)} & \textbf{$-$0.15$^{**}$ (0.06)} & \textbf{$-$0.15$^{**}$ (0.06)} & \textbf{$-$0.15$^{*}$ (0.06)} \\ 
  Decision 5 & \textbf{0.20$^{**}$ (0.05)} & \textbf{0.22$^{**}$ (0.05)} & \textbf{0.22$^{**}$ (0.05)} & \textbf{0.22$^{**}$ (0.05)} \\ 
  Decision 9 & \textbf{0.13$^{**}$ (0.05)} & \textbf{0.14$^{**}$ (0.05)} & \textbf{0.14$^{**}$ (0.05)} & \textbf{0.14$^{**}$ (0.05)} \\ 
  4 Actions:Decision 5 & $-$0.10 (0.07) & $-$0.11 (0.07) & $-$0.11 (0.07) & $-$0.11 (0.07) \\ 
  6 Actions:Decision 5 & $-$0.04 (0.07) & $-$0.05 (0.07) & $-$0.05 (0.07) & $-$0.05 (0.07) \\ 
  4 Actions:Decision 9 & $-$0.00 (0.07) & 0.00 (0.07) & 0.00 (0.07) & 0.00 (0.07) \\ 
  6 Actions:Decision 9 & 0.13 (0.07) & \textbf{0.14$^{*}$ (0.07)} & \textbf{0.14$^{*}$ (0.07)} & \textbf{0.14$^{*}$ (0.07)} \\ 
 \hline \\[-1.8ex] 
Observations & 615 & 615 & 615 & 615 \\ 
\hline \\[-1.8ex] 
\multicolumn{5}{l} {\textit{Note:} The reference condition for Actions is 2 Actions; the reference condition for Decision}\\
\multicolumn{5}{l}{is Decision 1. (1)-(4) reflect that those are four different linear mixed models, (1) tests the } \\
\multicolumn{5}{l}{effect of perceived autonomy, (2) meaningfulness, (3) task motivation, and (4) task load on } \\
\multicolumn{5}{l}{the percentage of correct answers.$^{*}$p$<$0.05; $^{**}$p$<$0.01} \\
\end{tabular} 
\end{table}

\begin{table}[t] 
\small
\centering 
  \caption{Linear mixed models with the dependent variable decision time} 
  \label{tab:hlmDecisionTime} 
\begin{tabular}{@{\extracolsep{5pt}}lcccc} 
\hline \\[-1.8ex] 
 & \multicolumn{4}{c}{\textit{Dependent variable:}} \\ 
\cline{2-5} 
\\[-1.8ex] & \multicolumn{4}{c}{MeanDecisionTime} \\ 
\\[-1.8ex] & (1) & (2) & (3) & (4)\\ 
\hline \\[-1.8ex] 
 Intercept & \textbf{4,579.66$^{**}$ (231.09)} & \textbf{4,119.91$^{**}$ (224.38)} & \textbf{3,937.71$^{**}$ (219.45)} & \textbf{4,646.55$^{**}$ (223.84)} \\ 
  Autonomy & $-$20.59 (36.06) & - & - & - \\ 
  Meaningfulness & - & \textbf{71.87$^{*}$ (31.19)} & - & - \\ 
  TaskMotivation & - & - & \textbf{103.36$^{**}$ (28.74)} & - \\ 
  TaskLoad & - & - & - & $-$47.87 (46.26) \\ 
  4 Actions & $-$571.84$^{*}$ \textbf{(227.14)} & \textbf{$-$561.84$^{*}$ (225.88)} & \textbf{$-$546.34$^{*}$ (224.71)} & \textbf{$-$560.91$^{*}$ (227.10)} \\ 
  6 Actions & $-$413.33 (227.18) & $-$430.06 (225.93) & $-$401.48 (224.65) & $-$407.41 (227.08) \\ 
  Decision 5 & \textbf{$-$1,248.90$^{**}$ (209.45)} & \textbf{$-$1,275.79$^{**}$ (209.56)} & \textbf{$-$1,282.57$^{**}$ (209.07)} & \textbf{$-$1,263.46$^{**}$ (208.92)} \\ 
  Decision 9 & \textbf{$-$850.22$^{**}$ (208.88)} & \textbf{$-$861.54$^{**}$ (209.45)} & \textbf{$-$851.25$^{**}$ (208.97)} & \textbf{$-$859.80$^{**}$ (208.94)} \\ 
  Decision 10 & \textbf{$-$982.27$^{**}$ (208.96)} & \textbf{$-$983.29$^{**}$ (209.44)} & \textbf{$-$951.58$^{**}$ (209.09)} & \textbf{$-$978.43$^{**}$ (208.87)} \\ 
  4 Actions:Decision 5 & 569.88 (296.45) & 578.33 (297.26) & \textbf{589.73$^{*}$ (296.64)} & 573.88 (296.47) \\ 
  6 Actions:Decision 5 & 442.64 (296.39) & 457.13 (297.30) & 453.53 (296.62) & 447.85 (296.47) \\ 
  4 Actions:Decision 9 & 421.05 (296.34) & 404.43 (297.33) & 387.45 (296.74) & 417.49 (296.47) \\ 
  6 Actions:Decision 9 & 347.48 (296.34) & 341.12 (297.26) & 342.46 (296.61) & 347.95 (296.48) \\ 
  4 Actions:Decision 10 & 503.66 (296.89) & 480.36 (297.31) & 447.91 (296.88) & 491.59 (296.47) \\ 
  6 Actions:Decision 10 & 450.20 (296.51) & 442.54 (297.26) & 409.64 (296.77) & 444.20 (296.47) \\ 
 \hline \\[-1.8ex] 
Observations & 820 & 820 & 820 & 820 \\ 
\hline \\[-1.8ex] 
\multicolumn{5}{l} {\textit{Note:} The reference condition for Actions is 2 Actions; the reference condition for Decision is Decision 1. (1)-(4) reflect that those } \\
\multicolumn{5}{l}{are four different linear mixed models, (1) tests the effect of perceived autonomy, (2) meaningfulness, (3) task motivation, and } \\
\multicolumn{5}{l}{(4) task load on the percentage of correct answers.$^{*}$p$<$0.05; $^{**}$p$<$0.01} \\
\end{tabular} 
\end{table} 
\newpage

\subsection{Motivational variables and their relation to decision-making}
\label{sec:AnalysisMotivation}
Research Questions 2a-b asked whether perceived autonomy, meaningfulness and task motivation
affect decision-related variables. 
To explore this, we calculated another set of linear mixed models with the dependent variables decision accuracy (\autoref{tab:hlmDecisionAccuracy}) and decision time (\autoref{tab:hlmDecisionTime}). 
For these analyses, we excluded participants from condition \emph{One}, since there was virtually no variance between participants' decision accuracy in this condition given that they always received the successful solution as a recommendation by the AI-based system. We also excluded the final round of the study in which all participants were forced to take a decision leading to a crash, thus having no variance between participants in their decision accuracy. 

The linear mixed models thus included dummy-coded variables that modeled \emph{Two} \textit{Selectable Action} as the reference condition, dummy-coded variables that modeled \emph{Decision Round 1} as the reference condition, and interaction effects between \emph{Selectable Actions} and \emph{Decision Round}. We included autonomy, meaningfulness, and task motivation in different linear mixed models to test whether they affect the decision-related variables beyond the experimental condition, beyond the decision round, and beyond the interactive effects between the experimental condition and the decision round.


For decision accuracy, we found that perceived autonomy was significantly positively related to participants' decision accuracy. This was not the case for perceived meaningfulness and motivation.

For the decision time, we found that autonomy was not associated with decision time. Instead, we found that higher motivation and meaningfulness were associated with significantly longer decision times.

Regarding Research Questions 2, we can thus respond that, 
\begin{enumerate}[label={RQ2\alph*.}]
    \item the accuracy of decisions made in the drone handover task was \textbf{positively related} to perceived autonomy. 
    \item participants' decision time was \textbf{positively related} to their perceived task meaningfulness and motivation.
\end{enumerate}

\section{Discussion}

The goal of this study was to examine whether an AI-based system that restricts the available choices of users to different degrees affects decision-related variables, perceived autonomy, and motivational aspects of human-AI interaction. The main findings of this study are 
\begin{enumerate}
    \item that having a choice positively affected perceived autonomy and perceived meaningfulness of the task with effects growing stronger over time,
    \item that when having a choice, perceived autonomy was associated with higher decision accuracy, while meaningfulness and motivation were associated with longer decision times, and
    \item taking the choice away from the participants positively affected their decision accuracy, owing to the nature of our task and the highly accurate AI-based support.
\end{enumerate} 

Together these findings provide first evidence on the impact that AI-based  systems can have on workers psychological needs and motivation and how this might affect overall work performance. In the following, we discuss these results in more detail and point out limitations of our study and opportunities for future work.

\subsection{The effects of having a choice}


Restricting choices for our participants improved their decision accuracy, decreased their decision time, and reduced the task load. These findings are in line with research showing that highly accurate and reliable AI-based systems can improve joint human-AI decision accuracy \cite{eisbach_optimizing_2023}. The positive effect on decision accuracy is also in line with De Vreede et al's \cite{de_vreede_design_2021} finding that restricting users' choices can positively affect joint human-AI decision accuracy. 
This is positive news for the use of AI systems in decision-making tasks. In our study, this effect was most clearly reflected in the better decision accuracy of the participants who only had one available action compared to the other conditions. Eventually, these participants had no real "choice" than to approve the system recommendation or to wait and ignore the recommendation, which would have led to a drone crash. At first glance, this may sound like an unrealistic scenario --- why even have a human overseer in such situations? However, given emerging legal regulations, such as the European Union's AI act, demanding the involvement of human overseers for high-risk tasks~\cite{Sterz_2024, Green_2022} (which includes the operation of autonomous drones, cf.\ Art.\ 6 and Annex 1(20) of the AI Act), we believe the scenario of having a human overseer simply approving AI-based recommendations with a limited degree of choice is realistic, though legal certainty on this matter will only be reached in the future. Importantly, those positive effects of restricting the \textit{Selectable Actions} on decision accuracy strongly depend on the reliability of the AI-based recommendations given that it is to expect that overseers will tend to follow also inaccurate AI-based recommendations~\cite{Green_2022, Parasuraman_2010}. Thus, the positive effects of restricting people's choices through AI-based recommendations will be less clear in cases where the AI-based system does not always offer the correct solution. Given that perfect and near-perfect recommendations by AI-based systems are unusual and will probably also be unusual in the forseeable future, this is an important consideration when restricting users' autonomy through AI-based systems.


The positive effects of restricting participants' choices on decision accuracy stand in contrast to their motivational effects. In contrast to only having one selectable action, having even just some choice (i.e., \emph{Two} \textit{Selectable Actions}) positively affected participants' perceived autonomy and meaningfulness of their task. Crucially, the differences between \emph{One} and other conditions became stronger over time. Consistent with the Self-Determination theory~\cite{Ryan_Deci_SDT_2000}, the work design literature~\cite{Hackman_oldham_JDS_1975, Morgeson_2003}, and the SMART model of work design~\cite{Parker_SMART_2023}, offering users some choices in their interaction with an AI-based system can be positively related to their need for autonomy and can have beneficial motivational effects reflected in their perceived meaningfulness of the task. In contrast, when users have little choice than to follow the recommendation of an AI-based system, over time they may start feeling useless as their behavior in the task is void of perceived meaning. In the long term, ignoring such detrimental effects on human needs and motivational variables may not only be negative for users motivation but also for their well-being with possible negative effects of jobs with little motivating appeal on people's mental health \cite{Parker_2014_beyondmotivation}. 

Consistent with this interpretation, whereas the number of selectable choices did not affect participants' motivation (and task satisfaction) in this study, we may expect that such effects could show during long term interaction with the system. Based on the SMART model and the broader work design literature, perceived meaningfulness is one of the strongest predictors of worker motivation and satisfaction over time~\cite{Humphrey_2007, Parker_2014_beyondmotivation, Parker_SMART_2023}. In our study, participants only experienced 10 decision rounds and an effective interaction with a system for about 10 minutes. This may not be enough to find long-term effects on user motivation. But even with just 10 minutes of interacting with the system, our participants started to question the meaningfulness of their role in this task. Thus, one key implication of the current study is that it will be crucial to conduct more long-term interaction studies to test for motivational effects in human-AI collaboration -- studies that are still rare in human-AI research \cite{Dong_2023}. Additionally, the effects on perceived autonomy and meaningfulness are important insights for the future design of human oversight jobs and of jobs where AI-based systems affect users' autonomy.

What may also be important to consider for future research is to find the right balance between considering users' need for autonomy and overwhelming users with too many choices. De Vreede et al. \cite{de_vreede_design_2021} showed that too much autonomy i.e., letting people decide freely when to use and when not to use an AI-based system, can be detrimental for joint human-AI decision accuracy. In our study, having \emph{Four} and \emph{Six} available choices led to higher perceived task load and both conditions were related to lower decision accuracy than having one available option. Eventually, having too much choice could even undermine users' perceived autonomy in a way that users will feel that they are not in a good position to make informed decisions. This curvilinear effect of autonomy is also reflected in evidence from the work design literature that shows that it is important to find a suitable degree of autonomy because autonomy misfit (i.e., too low and too much autonomy) can undermine people's motivation and well-being at work \cite{Stiglbauer_2018}, and can lead to detrimental effects such as users manipulating input data for AI-based systems to regain their autonomy \cite{Strich_2021}. 

In that regard, one implication of our study could be that as long as users have some choice, increasing available choices even more may not have further beneficial effects on variables such as perceived autonomy or meaningfulness. To put it differently, a certain degree of restriction in a way that an AI-based system restricts ineffective actions may not have detrimental effects on motivational variables important for effective long-term human-AI collaboration. In safety critical scenarios where it is important to minimize erroneous decisions, this is positive news. It may be possible to restrict users' choices in human-AI interactions in a way that leads to good joint performance without detrimentally impacting users' perceived autonomy --- as long as they are left with some choice. Clearly, this conclusion comes with ethical issues. For example, knowing that it is enough to ensure that people have some choice in order to uphold their perceived autonomy means that it may be enough to provide users with pseudo-choices or with the ability to slightly change the system with effects that are of little practical importance \cite{Dietvorst_2018}. Responsible design of human-AI collaboration needs to consider such potentially unethical ways of making people feel autonomous when all they have are pseudo-choices.


\subsection{Perceived autonomy is positively related to decision accuracy - meaningfulness and motivation are positively related to decision time}

In the conditions with some choice, those participants who felt more autonomous were also the ones who performed better. This supports propositions and findings by the work design literature that perceived autonomy can contribute to task performance~\cite{Humphrey_2007,Parker_SMART_2023}. In theory this kind of a positive effect may have been expected to also correspond to a stronger motivation of users. Yet, motivation was not significantly positively associated with decision accuracy in our analyses. We propose that this again calls for future research that examines the effects of design decisions that affect motivational aspects of human-AI collaboration in the long term. Potentially, in longer-term interactions it will be possible to provide insights on the expected relations between perceived autonomy, meaningfulness, and task motivation.

Interestingly, we found that task meaningfulness and motivation was related to a longer time that participants took for their decisions. On the one hand, this could mean that those participants who found the task to be more meaningful and who were motivated did their best in the task trying to think through their decisions for the \textit{Selectable Actions}. 
On the other hand, this association could mean that those people who found the task less meaningful and were less motivated wanted to finish it more quickly. Both of these interpretations for the findings would have been stronger if we had found relations between meaningfulness and/or motivation and decision accuracy that went beyond the effect of the experimental manipulations. Future research needs to more closely examine the relation between motivational variables and decision-related variables to uncover relations that are likely reciprocal: motivational variables affect decision-related variables (e.g., high motivation leads to high decision accuracy), and decision-related variables affect motivational variables (e.g., high decision accuracy leads to high motivation) \cite{Tay_2023}.




%

\subsection{Limitations and Future Work}

There are three main limitations of the current study that we want to highlight. First, participants only experienced a simulated drone oversight task without any real-world consequences. This can affect self-report items relating to users' perception of the oversight task. Nevertheless, we tried to simulate a realistic situation by showing participants videos of drone flights and by simulating time pressure. 

Second, participants only experienced ten scenarios and all of them ended in critical situations. In reality, people oversee drones for a longer period of time and optimally experience only a low number of critical situations. This may also be one reason for not finding significant differences between the experimental conditions for variables such as task motivation that may only play out after a prolonged period of time. Thus, for more generalizable insights on the role of selectable options on long-term implications on variables important for job satisfaction, future work would optimally simulate longer-term interactions with AI-based systems. However, this comes with many additional complexities and costs. We believe that for future research it will be worth investing these efforts given the findings and implications of the current study and given other recent research that highlights the importance of understanding the impact of human-AI interaction design on motivational variables and human needs~\cite{bucinca2024_intrinsicmotivation}.

Third, participants presumably had limited experience in flying drones and were no experts in human oversight tasks. This lack of expertise might have affected their overall perceived autonomy, as well as further motivational variables that we captured in this study. Still, we believe that the current study included a task where our participants were able to immerse themselves into and where they were capable of quickly understanding their responsibilities in the task. We are not sure how conducting the current study with experts would change the current results. This may depend on whether the person tasked with overseeing the drone was an expert drone pilot or an expert overseer. For expert drone pilots, reducing the autonomy may be more impactful than for expert overseers who may be used to having only limited points in time where they are supposed to intervene. However, this is a hypothesis for future work where it may be interesting to more closely examine motivational differences between people with different backgrounds who are supposed to fill the role of human overseers in high-risk tasks.

\section{Conclusion}

Beyond optimizing human-AI interactions for cognitive variables, it is important to also consider motivational variables which are often associated with satisfying human psychological needs \cite{Parker_SMART_2023, Ryan_Deci_SDT_2000}. In this regard, the current study examined the impact of restricting people's choices on their perceived autonomy and on downstream motivational variables and task performance via an AI-based decision support in an oversight task. Our findings highlight that restricting choices can be effective in helping people make accurate decisions. At the same time, restricting choices can undermine human overseers' perceived autonomy and important motivational variables such as the perceived meaningfulness of a task, particularly in the long term. The finding that people who felt more autonomous were also the ones who showed better decision accuracy further highlights the potential positive effects that considering human needs in human-AI interaction design can have on the joint performance of humans and AI-based systems. The most promising direction for future research that we see based on the current study is to conduct studies where humans and systems interact over the long term because this will help to better understand the impact of motivational variables in human-AI interactions that may only become stronger over time.

\begin{acks}
This work was funded by DFG grant 389792660 as part of TRR~248 -- CPEC, see \url{https://perspicuous-computing.science}
We thank Emilia Ellsiepen for all the support regarding the study design and execution.
\end{acks}

\bibliographystyle{ACM-Reference-Format}
\bibliography{bibliography}


\begin{thebibliography}{95}


\ifx \showCODEN    \undefined \def \showCODEN     #1{\unskip}     \fi
\ifx \showDOI      \undefined \def \showDOI       #1{#1}\fi
\ifx \showISBNx    \undefined \def \showISBNx     #1{\unskip}     \fi
\ifx \showISBNxiii \undefined \def \showISBNxiii  #1{\unskip}     \fi
\ifx \showISSN     \undefined \def \showISSN      #1{\unskip}     \fi
\ifx \showLCCN     \undefined \def \showLCCN      #1{\unskip}     \fi
\ifx \shownote     \undefined \def \shownote      #1{#1}          \fi
\ifx \showarticletitle \undefined \def \showarticletitle #1{#1}   \fi
\ifx \showURL      \undefined \def \showURL       {\relax}        \fi
\providecommand\bibfield[2]{#2}
\providecommand\bibinfo[2]{#2}
\providecommand\natexlab[1]{#1}
\providecommand\showeprint[2][]{arXiv:#2}

\bibitem[Alufaisan et~al\mbox{.}(2021)]%
        {alufaisan_does_2021}
\bibfield{author}{\bibinfo{person}{Yasmeen Alufaisan}, \bibinfo{person}{Laura~R. Marusich}, \bibinfo{person}{Jonathan~Z. Bakdash}, \bibinfo{person}{Yan Zhou}, {and} \bibinfo{person}{Murat Kantarcioglu}.} \bibinfo{year}{2021}\natexlab{}.
\newblock \showarticletitle{Does {Explainable} {Artificial} {Intelligence} {Improve} {Human} {Decision}-{Making}?}
\newblock \bibinfo{journal}{\emph{Proceedings of the AAAI Conference on Artificial Intelligence}} \bibinfo{volume}{35}, \bibinfo{number}{8} (\bibinfo{date}{May} \bibinfo{year}{2021}), \bibinfo{pages}{6618--6626}.
\newblock
\showISSN{2374-3468}
\urldef\tempurl%
\url{https://doi.org/10.1609/aaai.v35i8.16819}
\showDOI{\tempurl}
\newblock
\shownote{Number: 8}.


\bibitem[Ames and Fiske(2013)]%
        {ames_intentional_2013}
\bibfield{author}{\bibinfo{person}{Daniel~L. Ames} {and} \bibinfo{person}{Susan~T. Fiske}.} \bibinfo{year}{2013}\natexlab{}.
\newblock \showarticletitle{Intentional {Harms} {Are} {Worse}, {Even} {When} {They}’re {Not}}.
\newblock \bibinfo{journal}{\emph{Psychological Science}} \bibinfo{volume}{24}, \bibinfo{number}{9} (\bibinfo{date}{Sept.} \bibinfo{year}{2013}), \bibinfo{pages}{1755--1762}.
\newblock
\showISSN{0956-7976}
\urldef\tempurl%
\url{https://doi.org/10.1177/0956797613480507}
\showDOI{\tempurl}
\newblock
\shownote{Publisher: SAGE Publications Inc}.


\bibitem[Appelganc et~al\mbox{.}(2022)]%
        {appelganc_how_2022}
\bibfield{author}{\bibinfo{person}{Ksenia Appelganc}, \bibinfo{person}{Tobias Rieger}, \bibinfo{person}{Eileen Roesler}, {and} \bibinfo{person}{Dietrich Manzey}.} \bibinfo{year}{2022}\natexlab{}.
\newblock \showarticletitle{How {Much} {Reliability} {Is} {Enough}? {A} {Context}-{Specific} {View} on {Human} {Interaction} {With} ({Artificial}) {Agents} {From} {Different} {Perspectives}}.
\newblock \bibinfo{journal}{\emph{Journal of Cognitive Engineering and Decision Making}} \bibinfo{volume}{16}, \bibinfo{number}{4} (\bibinfo{date}{Dec.} \bibinfo{year}{2022}), \bibinfo{pages}{207--221}.
\newblock
\showISSN{1555-3434}
\urldef\tempurl%
\url{https://doi.org/10.1177/15553434221104615}
\showDOI{\tempurl}
\newblock
\shownote{Publisher: SAGE Publications}.


\bibitem[Bahner et~al\mbox{.}(2008)]%
        {Bahner_2008}
\bibfield{author}{\bibinfo{person}{J.~Elin Bahner}, \bibinfo{person}{Anke-Dorothea Hüper}, {and} \bibinfo{person}{Dietrich Manzey}.} \bibinfo{year}{2008}\natexlab{}.
\newblock \showarticletitle{Misuse of automated decision aids: Complacency, automation bias and the impact of training experience}.
\newblock \bibinfo{journal}{\emph{International Journal of Human-Computer Studies}} \bibinfo{volume}{66}, \bibinfo{number}{9} (\bibinfo{date}{Sept.} \bibinfo{year}{2008}), \bibinfo{pages}{688–699}.
\newblock
\showISSN{1071-5819}
\urldef\tempurl%
\url{https://doi.org/10.1016/j.ijhcs.2008.06.001}
\showDOI{\tempurl}


\bibitem[Bansal et~al\mbox{.}(2021)]%
        {bansal_does_2021}
\bibfield{author}{\bibinfo{person}{Gagan Bansal}, \bibinfo{person}{Tongshuang Wu}, \bibinfo{person}{Joyce Zhou}, \bibinfo{person}{Raymond Fok}, \bibinfo{person}{Besmira Nushi}, \bibinfo{person}{Ece Kamar}, \bibinfo{person}{Marco~Tulio Ribeiro}, {and} \bibinfo{person}{Daniel Weld}.} \bibinfo{year}{2021}\natexlab{}.
\newblock \showarticletitle{Does the {Whole} {Exceed} its {Parts}? {The} {Effect} of {AI} {Explanations} on {Complementary} {Team} {Performance}}. In \bibinfo{booktitle}{\emph{Proceedings of the 2021 {CHI} {Conference} on {Human} {Factors} in {Computing} {Systems}}} \emph{(\bibinfo{series}{{CHI} '21})}. \bibinfo{publisher}{Association for Computing Machinery}, \bibinfo{address}{New York, NY, USA}, \bibinfo{pages}{1--16}.
\newblock
\showISBNx{978-1-4503-8096-6}
\urldef\tempurl%
\url{https://doi.org/10.1145/3411764.3445717}
\showDOI{\tempurl}


\bibitem[Bu\c{c}inca(2024)]%
        {bucinca2024_intrinsicmotivation}
\bibfield{author}{\bibinfo{person}{Zana Bu\c{c}inca}.} \bibinfo{year}{2024}\natexlab{}.
\newblock \showarticletitle{Optimizing Decision-Maker's Intrinsic Motivation for Effective Human-AI Decision-Making}. In \bibinfo{booktitle}{\emph{Extended Abstracts of the 2024 CHI Conference on Human Factors in Computing Systems}} \emph{(\bibinfo{series}{CHI EA '24})}. \bibinfo{publisher}{Association for Computing Machinery}, \bibinfo{address}{New York, NY, USA}, Article \bibinfo{articleno}{439}, \bibinfo{numpages}{5}~pages.
\newblock
\showISBNx{9798400703317}
\urldef\tempurl%
\url{https://doi.org/10.1145/3613905.3638179}
\showDOI{\tempurl}


\bibitem[Buchanan and Kock(2001)]%
        {fandel_information_2001}
\bibfield{author}{\bibinfo{person}{John Buchanan} {and} \bibinfo{person}{Ned Kock}.} \bibinfo{year}{2001}\natexlab{}.
\newblock \showarticletitle{Information {Overload}: {A} {Decision} {Making} {Perspective}}.
\newblock In \bibinfo{booktitle}{\emph{Multiple {Criteria} {Decision} {Making} in the {New} {Millennium}}}, \bibfield{editor}{\bibinfo{person}{G.~Fandel}, \bibinfo{person}{W.~Trockel}, \bibinfo{person}{C.~D. Aliprantis}, \bibinfo{person}{Dan Kovenock}, \bibinfo{person}{Murat Köksalan}, {and} \bibinfo{person}{Stanley Zionts}} (Eds.). Vol.~\bibinfo{volume}{507}. \bibinfo{publisher}{Springer Berlin Heidelberg}, \bibinfo{address}{Berlin, Heidelberg}, \bibinfo{pages}{49--58}.
\newblock
\showISBNx{978-3-540-42377-5 978-3-642-56680-6}
\urldef\tempurl%
\url{https://doi.org/10.1007/978-3-642-56680-6_4}
\showDOI{\tempurl}
\newblock
\shownote{Series Title: Lecture Notes in Economics and Mathematical Systems}.


\bibitem[Bu{\c{c}}inca et~al\mbox{.}(2024)]%
        {buccinca2024towards}
\bibfield{author}{\bibinfo{person}{Zana Bu{\c{c}}inca}, \bibinfo{person}{Siddharth Swaroop}, \bibinfo{person}{Amanda~E Paluch}, \bibinfo{person}{Susan~A Murphy}, {and} \bibinfo{person}{Krzysztof~Z Gajos}.} \bibinfo{year}{2024}\natexlab{}.
\newblock \showarticletitle{Towards Optimizing Human-Centric Objectives in AI-Assisted Decision-Making With Offline Reinforcement Learning}.
\newblock \bibinfo{journal}{\emph{arXiv preprint arXiv:2403.05911}} (\bibinfo{year}{2024}).
\newblock


\bibitem[Buçinca et~al\mbox{.}(2020)]%
        {bucinca_proxy_2020}
\bibfield{author}{\bibinfo{person}{Zana Buçinca}, \bibinfo{person}{Phoebe Lin}, \bibinfo{person}{Krzysztof~Z. Gajos}, {and} \bibinfo{person}{Elena~L. Glassman}.} \bibinfo{year}{2020}\natexlab{}.
\newblock \showarticletitle{Proxy tasks and subjective measures can be misleading in evaluating explainable {AI} systems}. In \bibinfo{booktitle}{\emph{Proceedings of the 25th {International} {Conference} on {Intelligent} {User} {Interfaces}}}. \bibinfo{publisher}{ACM}, \bibinfo{address}{Cagliari Italy}, \bibinfo{pages}{454--464}.
\newblock
\showISBNx{978-1-4503-7118-6}
\urldef\tempurl%
\url{https://doi.org/10.1145/3377325.3377498}
\showDOI{\tempurl}


\bibitem[Buçinca et~al\mbox{.}(2021)]%
        {bucinca_trust_2021}
\bibfield{author}{\bibinfo{person}{Zana Buçinca}, \bibinfo{person}{Maja~Barbara Malaya}, {and} \bibinfo{person}{Krzysztof~Z. Gajos}.} \bibinfo{year}{2021}\natexlab{}.
\newblock \showarticletitle{To {Trust} or to {Think}: {Cognitive} {Forcing} {Functions} {Can} {Reduce} {Overreliance} on {AI} in {AI}-assisted {Decision}-making}.
\newblock \bibinfo{journal}{\emph{Proceedings of the ACM on Human-Computer Interaction}} \bibinfo{volume}{5}, \bibinfo{number}{CSCW1} (\bibinfo{date}{April} \bibinfo{year}{2021}), \bibinfo{pages}{1--21}.
\newblock
\showISSN{2573-0142}
\urldef\tempurl%
\url{https://doi.org/10.1145/3449287}
\showDOI{\tempurl}


\bibitem[Cabitza et~al\mbox{.}(2023)]%
        {cabitza_painting_2023}
\bibfield{author}{\bibinfo{person}{Federico Cabitza}, \bibinfo{person}{Andrea Campagner}, \bibinfo{person}{Chiara Natali}, \bibinfo{person}{Enea Parimbelli}, \bibinfo{person}{Luca Ronzio}, {and} \bibinfo{person}{Matteo Cameli}.} \bibinfo{year}{2023}\natexlab{}.
\newblock \showarticletitle{Painting the {Black} {Box} {White}: {Experimental} {Findings} from {Applying} {XAI} to an {ECG} {Reading} {Setting}}.
\newblock \bibinfo{journal}{\emph{Machine Learning and Knowledge Extraction}} \bibinfo{volume}{5}, \bibinfo{number}{1} (\bibinfo{date}{March} \bibinfo{year}{2023}), \bibinfo{pages}{269--286}.
\newblock
\showISSN{2504-4990}
\urldef\tempurl%
\url{https://doi.org/10.3390/make5010017}
\showDOI{\tempurl}
\newblock
\shownote{Number: 1 Publisher: Multidisciplinary Digital Publishing Institute}.


\bibitem[Calisto et~al\mbox{.}(2022)]%
        {calisto_breastscreening-ai_2022}
\bibfield{author}{\bibinfo{person}{Francisco~Maria Calisto}, \bibinfo{person}{Carlos Santiago}, \bibinfo{person}{Nuno Nunes}, {and} \bibinfo{person}{Jacinto~C. Nascimento}.} \bibinfo{year}{2022}\natexlab{}.
\newblock \showarticletitle{{BreastScreening}-{AI}: {Evaluating} medical intelligent agents for human-{AI} interactions}.
\newblock \bibinfo{journal}{\emph{Artificial Intelligence in Medicine}}  \bibinfo{volume}{127} (\bibinfo{date}{May} \bibinfo{year}{2022}), \bibinfo{pages}{102285}.
\newblock
\showISSN{1873-2860}
\urldef\tempurl%
\url{https://doi.org/10.1016/j.artmed.2022.102285}
\showDOI{\tempurl}


\bibitem[Campbell et~al\mbox{.}(1993)]%
        {campbell1993theory}
\bibfield{author}{\bibinfo{person}{J.P. Campbell}, \bibinfo{person}{R.A. McCloy}, \bibinfo{person}{S.H. Oppler}, {and} \bibinfo{person}{C.E. Sager}.} \bibinfo{year}{1993}\natexlab{}.
\newblock \showarticletitle{A Theory of Performance}.
\newblock In \bibinfo{booktitle}{\emph{Personnel Selection in Organizations}}, \bibfield{editor}{\bibinfo{person}{N.~Schmitt} {and} \bibinfo{person}{W.C. Borman}} (Eds.). \bibinfo{publisher}{Jossey-Bass Publishers}, \bibinfo{address}{San Francisco}, \bibinfo{pages}{35--70}.
\newblock


\bibitem[Cau et~al\mbox{.}(2023)]%
        {cau_supporting_2023}
\bibfield{author}{\bibinfo{person}{Federico~Maria Cau}, \bibinfo{person}{Hanna Hauptmann}, \bibinfo{person}{Lucio~Davide Spano}, {and} \bibinfo{person}{Nava Tintarev}.} \bibinfo{year}{2023}\natexlab{}.
\newblock \showarticletitle{Supporting {High}-{Uncertainty} {Decisions} through {AI} and {Logic}-{Style} {Explanations}}. In \bibinfo{booktitle}{\emph{Proceedings of the 28th {International} {Conference} on {Intelligent} {User} {Interfaces}}} \emph{(\bibinfo{series}{{IUI} '23})}. \bibinfo{publisher}{Association for Computing Machinery}, \bibinfo{address}{New York, NY, USA}, \bibinfo{pages}{251--263}.
\newblock
\showISBNx{9798400701061}
\urldef\tempurl%
\url{https://doi.org/10.1145/3581641.3584080}
\showDOI{\tempurl}


\bibitem[Cerasoli et~al\mbox{.}(2016)]%
        {Cerasoli_Meta_2016}
\bibfield{author}{\bibinfo{person}{Christopher~P. Cerasoli}, \bibinfo{person}{Jessica~M. Nicklin}, {and} \bibinfo{person}{Alexander~S. Nassrelgrgawi}.} \bibinfo{year}{2016}\natexlab{}.
\newblock \showarticletitle{Performance, incentives, and needs for autonomy, competence, and relatedness: a meta-analysis}.
\newblock \bibinfo{journal}{\emph{Motivation and Emotion}} \bibinfo{volume}{40}, \bibinfo{number}{6} (\bibinfo{date}{Sept.} \bibinfo{year}{2016}), \bibinfo{pages}{781–813}.
\newblock
\showISSN{1573-6644}
\urldef\tempurl%
\url{https://doi.org/10.1007/s11031-016-9578-2}
\showDOI{\tempurl}


\bibitem[Chernev et~al\mbox{.}(2015)]%
        {chernev_choice_2015}
\bibfield{author}{\bibinfo{person}{Alexander Chernev}, \bibinfo{person}{Ulf Böckenholt}, {and} \bibinfo{person}{Joseph Goodman}.} \bibinfo{year}{2015}\natexlab{}.
\newblock \showarticletitle{Choice overload: {A} conceptual review and meta-analysis}.
\newblock \bibinfo{journal}{\emph{Journal of Consumer Psychology}} \bibinfo{volume}{25}, \bibinfo{number}{2} (\bibinfo{year}{2015}), \bibinfo{pages}{333--358}.
\newblock
\showISSN{1532-7663}
\urldef\tempurl%
\url{https://doi.org/10.1016/j.jcps.2014.08.002}
\showDOI{\tempurl}
\newblock
\shownote{\_eprint: https://onlinelibrary.wiley.com/doi/pdf/10.1016/j.jcps.2014.08.002}.


\bibitem[Das et~al\mbox{.}(2024)]%
        {Das_2024}
\bibfield{author}{\bibinfo{person}{Anwesha Das}, \bibinfo{person}{Zekun Wu}, \bibinfo{person}{Iza Skrjanec}, {and} \bibinfo{person}{Anna~Maria Feit}.} \bibinfo{year}{2024}\natexlab{}.
\newblock \showarticletitle{Shifting Focus with HCEye: Exploring the Dynamics of Visual Highlighting and Cognitive Load on User Attention and Saliency Prediction}.
\newblock \bibinfo{journal}{\emph{Proceedings of the ACM on Human-Computer Interaction}} \bibinfo{volume}{8}, \bibinfo{number}{ETRA} (\bibinfo{date}{May} \bibinfo{year}{2024}), \bibinfo{pages}{1–18}.
\newblock
\showISSN{2573-0142}
\urldef\tempurl%
\url{https://doi.org/10.1145/3655610}
\showDOI{\tempurl}


\bibitem[De~Vreede et~al\mbox{.}(2021)]%
        {de_vreede_design_2021}
\bibfield{author}{\bibinfo{person}{Triparna De~Vreede}, \bibinfo{person}{Mukhunth Raghavan}, {and} \bibinfo{person}{Gert-Jan De~Vreede}.} \bibinfo{year}{2021}\natexlab{}.
\newblock \showarticletitle{Design {Foundations} for {AI} {Assisted} {Decision} {Making}: {A} {Self} {Determination} {Theory} {Approach}}. In \bibinfo{booktitle}{\emph{Proceedings of the 54th Annual Hawaii International Conference on System Sciences}}.
\newblock
\urldef\tempurl%
\url{https://doi.org/10.24251/HICSS.2021.019}
\showDOI{\tempurl}


\bibitem[Demerouti and Bakker(2008)]%
        {demerouti2008oldenburg}
\bibfield{author}{\bibinfo{person}{Evangelia Demerouti} {and} \bibinfo{person}{Arnold~B Bakker}.} \bibinfo{year}{2008}\natexlab{}.
\newblock \showarticletitle{The Oldenburg Burnout Inventory: A good alternative to measure burnout and engagement}.
\newblock \bibinfo{journal}{\emph{Handbook of stress and burnout in health care}} \bibinfo{volume}{65}, \bibinfo{number}{7} (\bibinfo{year}{2008}), \bibinfo{pages}{1--25}.
\newblock


\bibitem[Dietvorst et~al\mbox{.}(2018)]%
        {Dietvorst_2018}
\bibfield{author}{\bibinfo{person}{Berkeley~J. Dietvorst}, \bibinfo{person}{Joseph~P. Simmons}, {and} \bibinfo{person}{Cade Massey}.} \bibinfo{year}{2018}\natexlab{}.
\newblock \showarticletitle{Overcoming Algorithm Aversion: People Will Use Imperfect Algorithms If They Can (Even Slightly) Modify Them}.
\newblock \bibinfo{journal}{\emph{Management Science}} \bibinfo{volume}{64}, \bibinfo{number}{3} (\bibinfo{date}{March} \bibinfo{year}{2018}), \bibinfo{pages}{1155–1170}.
\newblock
\showISSN{1526-5501}
\urldef\tempurl%
\url{https://doi.org/10.1287/mnsc.2016.2643}
\showDOI{\tempurl}


\bibitem[Dong et~al\mbox{.}(2023)]%
        {Dong_2023}
\bibfield{author}{\bibinfo{person}{Mengchen Dong}, \bibinfo{person}{Jean-Francois Bonnefon}, {and} \bibinfo{person}{Iyad Rahwan}.} \bibinfo{year}{2023}\natexlab{}.
\newblock \showarticletitle{Toward Human-Centered AI Management: Methodological Challenges and Future Directions}.
\newblock \bibinfo{journal}{\emph{SSRN Electronic Journal}} (\bibinfo{year}{2023}).
\newblock
\showISSN{1556-5068}
\urldef\tempurl%
\url{https://doi.org/10.2139/ssrn.4444322}
\showDOI{\tempurl}


\bibitem[Eisbach et~al\mbox{.}(2023)]%
        {eisbach_optimizing_2023}
\bibfield{author}{\bibinfo{person}{Simon Eisbach}, \bibinfo{person}{Markus Langer}, {and} \bibinfo{person}{Guido Hertel}.} \bibinfo{year}{2023}\natexlab{}.
\newblock \showarticletitle{Optimizing human-{AI} collaboration: {Effects} of motivation and accuracy information in {AI}-supported decision-making}.
\newblock \bibinfo{journal}{\emph{Computers in Human Behavior: Artificial Humans}} \bibinfo{volume}{1}, \bibinfo{number}{2} (\bibinfo{date}{Aug.} \bibinfo{year}{2023}), \bibinfo{pages}{100015}.
\newblock
\showISSN{2949-8821}
\urldef\tempurl%
\url{https://doi.org/10.1016/j.chbah.2023.100015}
\showDOI{\tempurl}


\bibitem[Faul et~al\mbox{.}(2007)]%
        {faul_gpower_2007}
\bibfield{author}{\bibinfo{person}{Franz Faul}, \bibinfo{person}{Edgar Erdfelder}, \bibinfo{person}{Albert-Georg Lang}, {and} \bibinfo{person}{Axel Buchner}.} \bibinfo{year}{2007}\natexlab{}.
\newblock \showarticletitle{G*{Power} 3: {A} flexible statistical power analysis program for the social, behavioral, and biomedical sciences}.
\newblock \bibinfo{journal}{\emph{Behavior Research Methods}} \bibinfo{volume}{39}, \bibinfo{number}{2} (\bibinfo{date}{May} \bibinfo{year}{2007}), \bibinfo{pages}{175--191}.
\newblock
\showISSN{1554-3528}
\urldef\tempurl%
\url{https://doi.org/10.3758/BF03193146}
\showDOI{\tempurl}


\bibitem[Feit et~al\mbox{.}(2020)]%
        {Feit2020}
\bibfield{author}{\bibinfo{person}{Anna~Maria Feit}, \bibinfo{person}{Lukas Vordemann}, \bibinfo{person}{Seonwook Park}, \bibinfo{person}{Caterina Berube}, {and} \bibinfo{person}{Otmar Hilliges}.} \bibinfo{year}{2020}\natexlab{}.
\newblock \showarticletitle{Detecting relevance during decision-making from eye movements for UI adaptation}.
\newblock \bibinfo{journal}{\emph{Eye Tracking Research and Applications Symposium (ETRA)}} (\bibinfo{date}{2} \bibinfo{year}{2020}).
\newblock
\showISBNx{9781450371339}
\urldef\tempurl%
\url{https://doi.org/10.1145/3379155.3391321}
\showDOI{\tempurl}


\bibitem[Gagné et~al\mbox{.}(2010)]%
        {gagne_motivation_2010}
\bibfield{author}{\bibinfo{person}{Marylène Gagné}, \bibinfo{person}{Jacques Forest}, \bibinfo{person}{Marie-Hélène Gilbert}, \bibinfo{person}{Caroline Aubé}, \bibinfo{person}{Estelle Morin}, {and} \bibinfo{person}{Angela Malorni}.} \bibinfo{year}{2010}\natexlab{}.
\newblock \showarticletitle{The {Motivation} at {Work} {Scale}: {Validation} {Evidence} in {Two} {Languages}}.
\newblock \bibinfo{journal}{\emph{Educational and Psychological Measurement}} \bibinfo{volume}{70}, \bibinfo{number}{4} (\bibinfo{date}{Aug.} \bibinfo{year}{2010}), \bibinfo{pages}{628--646}.
\newblock
\showISSN{0013-1644, 1552-3888}
\urldef\tempurl%
\url{https://doi.org/10.1177/0013164409355698}
\showDOI{\tempurl}


\bibitem[Gailey and Falk(2008)]%
        {gailey_attribution_2008}
\bibfield{author}{\bibinfo{person}{Jeannine~A. Gailey} {and} \bibinfo{person}{R.~Frank Falk}.} \bibinfo{year}{2008}\natexlab{}.
\newblock \showarticletitle{Attribution of {Responsibility} as a {Multidimensional} {Concept}}.
\newblock \bibinfo{journal}{\emph{Sociological Spectrum}} \bibinfo{volume}{28}, \bibinfo{number}{6} (\bibinfo{date}{Sept.} \bibinfo{year}{2008}), \bibinfo{pages}{659--680}.
\newblock
\showISSN{0273-2173}
\urldef\tempurl%
\url{https://doi.org/10.1080/02732170802342958}
\showDOI{\tempurl}
\newblock
\shownote{Publisher: Routledge \_eprint: https://doi.org/10.1080/02732170802342958}.


\bibitem[Green(2022)]%
        {Green_2022}
\bibfield{author}{\bibinfo{person}{Ben Green}.} \bibinfo{year}{2022}\natexlab{}.
\newblock \showarticletitle{The flaws of policies requiring human oversight of government algorithms}.
\newblock \bibinfo{journal}{\emph{Computer Law \& Security Review}}  \bibinfo{volume}{45} (\bibinfo{date}{July} \bibinfo{year}{2022}), \bibinfo{pages}{105681}.
\newblock
\showISSN{0267-3649}
\urldef\tempurl%
\url{https://doi.org/10.1016/j.clsr.2022.105681}
\showDOI{\tempurl}


\bibitem[Green and Chen(2019a)]%
        {Green_2019_disparate}
\bibfield{author}{\bibinfo{person}{Ben Green} {and} \bibinfo{person}{Yiling Chen}.} \bibinfo{year}{2019}\natexlab{a}.
\newblock \showarticletitle{Disparate Interactions: An Algorithm-in-the-Loop Analysis of Fairness in Risk Assessments}. In \bibinfo{booktitle}{\emph{Proceedings of the Conference on Fairness, Accountability, and Transparency}} \emph{(\bibinfo{series}{FAT* ’19})}. \bibinfo{publisher}{ACM}.
\newblock
\urldef\tempurl%
\url{https://doi.org/10.1145/3287560.3287563}
\showDOI{\tempurl}


\bibitem[Green and Chen(2019b)]%
        {green_principles_2019}
\bibfield{author}{\bibinfo{person}{Ben Green} {and} \bibinfo{person}{Yiling Chen}.} \bibinfo{year}{2019}\natexlab{b}.
\newblock \showarticletitle{The {Principles} and {Limits} of {Algorithm}-in-the-{Loop} {Decision} {Making}}.
\newblock \bibinfo{journal}{\emph{Proc. ACM Hum.-Comput. Interact.}} \bibinfo{volume}{3}, \bibinfo{number}{CSCW} (\bibinfo{date}{Nov.} \bibinfo{year}{2019}), \bibinfo{pages}{50:1--50:24}.
\newblock
\urldef\tempurl%
\url{https://doi.org/10.1145/3359152}
\showDOI{\tempurl}


\bibitem[Grgić-Hlača et~al\mbox{.}(2019)]%
        {grgic-hlaca_human_2019}
\bibfield{author}{\bibinfo{person}{Nina Grgić-Hlača}, \bibinfo{person}{Christoph Engel}, {and} \bibinfo{person}{Krishna~P. Gummadi}.} \bibinfo{year}{2019}\natexlab{}.
\newblock \showarticletitle{Human {Decision} {Making} with {Machine} {Assistance}: {An} {Experiment} on {Bailing} and {Jailing}}.
\newblock \bibinfo{journal}{\emph{Proceedings of the ACM on Human-Computer Interaction}} \bibinfo{volume}{3}, \bibinfo{number}{CSCW} (\bibinfo{date}{Nov.} \bibinfo{year}{2019}), \bibinfo{pages}{1--25}.
\newblock
\showISSN{2573-0142}
\urldef\tempurl%
\url{https://doi.org/10.1145/3359280}
\showDOI{\tempurl}


\bibitem[Gu et~al\mbox{.}(2023)]%
        {gu_improving_2023}
\bibfield{author}{\bibinfo{person}{Hongyan Gu}, \bibinfo{person}{Yuan Liang}, \bibinfo{person}{Yifan Xu}, \bibinfo{person}{Christopher~Kazu Williams}, \bibinfo{person}{Shino Magaki}, \bibinfo{person}{Negar Khanlou}, \bibinfo{person}{Harry Vinters}, \bibinfo{person}{Zesheng Chen}, \bibinfo{person}{Shuo Ni}, \bibinfo{person}{Chunxu Yang}, \bibinfo{person}{Wenzhong Yan}, \bibinfo{person}{Xinhai~Robert Zhang}, \bibinfo{person}{Yang Li}, \bibinfo{person}{Mohammad Haeri}, {and} \bibinfo{person}{Xiang~‘Anthony’ Chen}.} \bibinfo{year}{2023}\natexlab{}.
\newblock \showarticletitle{Improving {Workflow} {Integration} with {xPath}: {Design} and {Evaluation} of a {Human}-{AI} {Diagnosis} {System} in {Pathology}}.
\newblock \bibinfo{journal}{\emph{ACM Trans. Comput.-Hum. Interact.}} \bibinfo{volume}{30}, \bibinfo{number}{2} (\bibinfo{date}{March} \bibinfo{year}{2023}), \bibinfo{pages}{28:1--28:37}.
\newblock
\showISSN{1073-0516}
\urldef\tempurl%
\url{https://doi.org/10.1145/3577011}
\showDOI{\tempurl}


\bibitem[Gundappa et~al\mbox{.}(2024)]%
        {gundappa2024designinformativetakeoverrequests}
\bibfield{author}{\bibinfo{person}{Ashwini Gundappa}, \bibinfo{person}{Emilia Ellsiepen}, \bibinfo{person}{Lukas Schmitz}, \bibinfo{person}{Frederik Wiehr}, {and} \bibinfo{person}{Vera Demberg}.} \bibinfo{year}{2024}\natexlab{}.
\newblock \bibinfo{title}{The Design of Informative Take-Over Requests for Semi-Autonomous Cyber-Physical Systems: Combining Spoken Language and Visual Icons in a Drone-Controller Setting}.
\newblock
\newblock
\showeprint[arxiv]{2409.08253}~[cs.HC]
\urldef\tempurl%
\url{https://arxiv.org/abs/2409.08253}
\showURL{%
\tempurl}


\bibitem[Hackman(1980)]%
        {Hackman_redesign_1980}
\bibfield{author}{\bibinfo{person}{J.~Richard Hackman}.} \bibinfo{year}{1980}\natexlab{}.
\newblock \showarticletitle{Work redesign and motivation.}
\newblock \bibinfo{journal}{\emph{Professional Psychology}} \bibinfo{volume}{11}, \bibinfo{number}{3} (\bibinfo{date}{June} \bibinfo{year}{1980}), \bibinfo{pages}{445–455}.
\newblock
\showISSN{0033-0175}
\urldef\tempurl%
\url{https://doi.org/10.1037/0735-7028.11.3.445}
\showDOI{\tempurl}


\bibitem[Hackman and Oldham(1975)]%
        {Hackman_oldham_JDS_1975}
\bibfield{author}{\bibinfo{person}{J.~Richard Hackman} {and} \bibinfo{person}{Greg~R. Oldham}.} \bibinfo{year}{1975}\natexlab{}.
\newblock \showarticletitle{Development of the Job Diagnostic Survey.}
\newblock \bibinfo{journal}{\emph{Journal of Applied Psychology}} \bibinfo{volume}{60}, \bibinfo{number}{2} (\bibinfo{date}{April} \bibinfo{year}{1975}), \bibinfo{pages}{159–170}.
\newblock
\showISSN{0021-9010}
\urldef\tempurl%
\url{https://doi.org/10.1037/h0076546}
\showDOI{\tempurl}


\bibitem[Hart and Staveland(1988)]%
        {hart_development_1988}
\bibfield{author}{\bibinfo{person}{Sandra~G. Hart} {and} \bibinfo{person}{Lowell~E. Staveland}.} \bibinfo{year}{1988}\natexlab{}.
\newblock \showarticletitle{Development of {NASA}-{TLX} ({Task} {Load} {Index}): {Results} of {Empirical} and {Theoretical} {Research}}.
\newblock In \bibinfo{booktitle}{\emph{Advances in {Psychology}}}, \bibfield{editor}{\bibinfo{person}{Peter~A. Hancock} {and} \bibinfo{person}{Najmedin Meshkati}} (Eds.). \bibinfo{series}{Human {Mental} {Workload}}, Vol.~\bibinfo{volume}{52}. \bibinfo{publisher}{North-Holland}, \bibinfo{pages}{139--183}.
\newblock
\urldef\tempurl%
\url{https://doi.org/10.1016/S0166-4115(08)62386-9}
\showDOI{\tempurl}


\bibitem[Herm(2023)]%
        {herm_impact_2023}
\bibfield{author}{\bibinfo{person}{Lukas-Valentin Herm}.} \bibinfo{year}{2023}\natexlab{}.
\newblock \showarticletitle{{IMPACT} {OF} {EXPLAINABLE} {AI} {ON} {COGNITIVE} {LOAD}: {INSIGHTS} {FROM} {AN} {EMPIRICAL} {STUDY}}.
\newblock \bibinfo{journal}{\emph{ECIS 2023 Research Papers}} (\bibinfo{year}{2023}).
\newblock


\bibitem[Hinds et~al\mbox{.}(2004)]%
        {hinds_whose_2004}
\bibfield{author}{\bibinfo{person}{Pamela Hinds}, \bibinfo{person}{Teresa Roberts}, {and} \bibinfo{person}{Hank Jones}.} \bibinfo{year}{2004}\natexlab{}.
\newblock \showarticletitle{Whose {Job} {Is} {It} {Anyway}? {A} {Study} of {Human}-{Robot} {Interaction} in a {Collaborative} {Task}}.
\newblock \bibinfo{journal}{\emph{Human-Computer Interaction}} \bibinfo{volume}{19}, \bibinfo{number}{1} (\bibinfo{date}{June} \bibinfo{year}{2004}), \bibinfo{pages}{151--181}.
\newblock
\showISSN{0737-0024}
\urldef\tempurl%
\url{https://doi.org/10.1207/s15327051hci1901&2_7}
\showDOI{\tempurl}


\bibitem[Hoffman et~al\mbox{.}(2019)]%
        {hoffman_metrics_2019}
\bibfield{author}{\bibinfo{person}{Robert~R. Hoffman}, \bibinfo{person}{Shane~T. Mueller}, \bibinfo{person}{Gary Klein}, {and} \bibinfo{person}{Jordan Litman}.} \bibinfo{year}{2019}\natexlab{}.
\newblock \bibinfo{title}{Metrics for {Explainable} {AI}: {Challenges} and {Prospects}}.
\newblock
\newblock
\urldef\tempurl%
\url{https://doi.org/10.48550/arXiv.1812.04608}
\showDOI{\tempurl}
\newblock
\shownote{arXiv:1812.04608 [cs]}.


\bibitem[Hudon et~al\mbox{.}(2021)]%
        {hudon_explainable_2021}
\bibfield{author}{\bibinfo{person}{Antoine Hudon}, \bibinfo{person}{Théophile Demazure}, \bibinfo{person}{Alexander Karran}, \bibinfo{person}{Pierre-Majorique Léger}, {and} \bibinfo{person}{Sylvain Sénécal}.} \bibinfo{year}{2021}\natexlab{}.
\newblock \showarticletitle{Explainable {Artificial} {Intelligence} ({XAI}): {How} the {Visualization} of {AI} {Predictions} {Affects} {User} {Cognitive} {Load} and {Confidence}}. In \bibinfo{booktitle}{\emph{Information {Systems} and {Neuroscience}}}, \bibfield{editor}{\bibinfo{person}{Fred~D. Davis}, \bibinfo{person}{René Riedl}, \bibinfo{person}{Jan vom Brocke}, \bibinfo{person}{Pierre-Majorique Léger}, \bibinfo{person}{Adriane~B. Randolph}, {and} \bibinfo{person}{Gernot Müller-Putz}} (Eds.). \bibinfo{publisher}{Springer International Publishing}, \bibinfo{address}{Cham}, \bibinfo{pages}{237--246}.
\newblock
\showISBNx{978-3-030-88900-5}
\urldef\tempurl%
\url{https://doi.org/10.1007/978-3-030-88900-5_27}
\showDOI{\tempurl}


\bibitem[Humphrey et~al\mbox{.}(2007)]%
        {Humphrey_2007}
\bibfield{author}{\bibinfo{person}{Stephen~E. Humphrey}, \bibinfo{person}{Jennifer~D. Nahrgang}, {and} \bibinfo{person}{Frederick~P. Morgeson}.} \bibinfo{year}{2007}\natexlab{}.
\newblock \showarticletitle{Integrating motivational, social, and contextual work design features: A meta-analytic summary and theoretical extension of the work design literature.}
\newblock \bibinfo{journal}{\emph{Journal of Applied Psychology}} \bibinfo{volume}{92}, \bibinfo{number}{5} (\bibinfo{year}{2007}), \bibinfo{pages}{1332–1356}.
\newblock
\showISSN{0021-9010}
\urldef\tempurl%
\url{https://doi.org/10.1037/0021-9010.92.5.1332}
\showDOI{\tempurl}


\bibitem[Jacobs et~al\mbox{.}(2021)]%
        {jacobs_how_2021}
\bibfield{author}{\bibinfo{person}{Maia Jacobs}, \bibinfo{person}{Melanie~F. Pradier}, \bibinfo{person}{Thomas~H. McCoy}, \bibinfo{person}{Roy~H. Perlis}, \bibinfo{person}{Finale Doshi-Velez}, {and} \bibinfo{person}{Krzysztof~Z. Gajos}.} \bibinfo{year}{2021}\natexlab{}.
\newblock \showarticletitle{How machine-learning recommendations influence clinician treatment selections: the example of antidepressant selection}.
\newblock \bibinfo{journal}{\emph{Translational Psychiatry}} \bibinfo{volume}{11}, \bibinfo{number}{1} (\bibinfo{date}{Feb.} \bibinfo{year}{2021}), \bibinfo{pages}{1--9}.
\newblock
\showISSN{2158-3188}
\urldef\tempurl%
\url{https://doi.org/10.1038/s41398-021-01224-x}
\showDOI{\tempurl}
\newblock
\shownote{Publisher: Nature Publishing Group}.


\bibitem[Kahr et~al\mbox{.}(2023)]%
        {kahr_it_2023}
\bibfield{author}{\bibinfo{person}{Patricia~K. Kahr}, \bibinfo{person}{Gerrit Rooks}, \bibinfo{person}{Martijn~C. Willemsen}, {and} \bibinfo{person}{Chris~C.P. Snijders}.} \bibinfo{year}{2023}\natexlab{}.
\newblock \showarticletitle{It {Seems} {Smart}, but {It} {Acts} {Stupid}: {Development} of {Trust} in {AI} {Advice} in a {Repeated} {Legal} {Decision}-{Making} {Task}}. In \bibinfo{booktitle}{\emph{Proceedings of the 28th {International} {Conference} on {Intelligent} {User} {Interfaces}}}. \bibinfo{publisher}{ACM}, \bibinfo{address}{Sydney NSW Australia}, \bibinfo{pages}{528--539}.
\newblock
\showISBNx{9798400701061}
\urldef\tempurl%
\url{https://doi.org/10.1145/3581641.3584058}
\showDOI{\tempurl}


\bibitem[Körber(2019)]%
        {korber_theoretical_2019}
\bibfield{author}{\bibinfo{person}{Moritz Körber}.} \bibinfo{year}{2019}\natexlab{}.
\newblock \showarticletitle{Theoretical {Considerations} and {Development} of a {Questionnaire} to {Measure} {Trust} in {Automation}}. In \bibinfo{booktitle}{\emph{Proceedings of the 20th {Congress} of the {International} {Ergonomics} {Association} ({IEA} 2018)}}, \bibfield{editor}{\bibinfo{person}{Sebastiano Bagnara}, \bibinfo{person}{Riccardo Tartaglia}, \bibinfo{person}{Sara Albolino}, \bibinfo{person}{Thomas Alexander}, {and} \bibinfo{person}{Yushi Fujita}} (Eds.). \bibinfo{publisher}{Springer International Publishing}, \bibinfo{address}{Cham}, \bibinfo{pages}{13--30}.
\newblock
\showISBNx{978-3-319-96074-6}
\urldef\tempurl%
\url{https://doi.org/10.1007/978-3-319-96074-6_2}
\showDOI{\tempurl}


\bibitem[Lai et~al\mbox{.}(2023)]%
        {Lai_2023}
\bibfield{author}{\bibinfo{person}{Vivian Lai}, \bibinfo{person}{Chacha Chen}, \bibinfo{person}{Alison Smith-Renner}, \bibinfo{person}{Q.~Vera Liao}, {and} \bibinfo{person}{Chenhao Tan}.} \bibinfo{year}{2023}\natexlab{}.
\newblock \showarticletitle{Towards a Science of Human-AI Decision Making: An Overview of Design Space in Empirical Human-Subject Studies}. In \bibinfo{booktitle}{\emph{2023 ACM Conference on Fairness, Accountability, and Transparency}} \emph{(\bibinfo{series}{FAccT ’23})}. \bibinfo{publisher}{ACM}.
\newblock
\urldef\tempurl%
\url{https://doi.org/10.1145/3593013.3594087}
\showDOI{\tempurl}


\bibitem[Langer et~al\mbox{.}(2022)]%
        {Langer_2022_trust}
\bibfield{author}{\bibinfo{person}{Markus Langer}, \bibinfo{person}{Cornelius~J. König}, \bibinfo{person}{Caroline Back}, {and} \bibinfo{person}{Victoria Hemsing}.} \bibinfo{year}{2022}\natexlab{}.
\newblock \showarticletitle{Trust in Artificial Intelligence: Comparing Trust Processes Between Human and Automated Trustees in Light of Unfair Bias}.
\newblock \bibinfo{journal}{\emph{Journal of Business and Psychology}} \bibinfo{volume}{38}, \bibinfo{number}{3} (\bibinfo{date}{June} \bibinfo{year}{2022}), \bibinfo{pages}{493–508}.
\newblock
\showISSN{1573-353X}
\urldef\tempurl%
\url{https://doi.org/10.1007/s10869-022-09829-9}
\showDOI{\tempurl}


\bibitem[Langer et~al\mbox{.}(2020)]%
        {Langer_2020_changing}
\bibfield{author}{\bibinfo{person}{Markus Langer}, \bibinfo{person}{Cornelius~J. König}, {and} \bibinfo{person}{Vivien Busch}.} \bibinfo{year}{2020}\natexlab{}.
\newblock \showarticletitle{Changing the means of managerial work: effects of automated decision support systems on personnel selection tasks}.
\newblock \bibinfo{journal}{\emph{Journal of Business and Psychology}} \bibinfo{volume}{36}, \bibinfo{number}{5} (\bibinfo{date}{Sept.} \bibinfo{year}{2020}), \bibinfo{pages}{751–769}.
\newblock
\showISSN{1573-353X}
\urldef\tempurl%
\url{https://doi.org/10.1007/s10869-020-09711-6}
\showDOI{\tempurl}


\bibitem[Langer and Landers(2021)]%
        {LangerLanders_2021}
\bibfield{author}{\bibinfo{person}{Markus Langer} {and} \bibinfo{person}{Richard~N. Landers}.} \bibinfo{year}{2021}\natexlab{}.
\newblock \showarticletitle{The future of artificial intelligence at work: A review on effects of decision automation and augmentation on workers targeted by algorithms and third-party observers}.
\newblock \bibinfo{journal}{\emph{Computers in Human Behavior}}  \bibinfo{volume}{123} (\bibinfo{date}{Oct.} \bibinfo{year}{2021}), \bibinfo{pages}{106878}.
\newblock
\showISSN{0747-5632}
\urldef\tempurl%
\url{https://doi.org/10.1016/j.chb.2021.106878}
\showDOI{\tempurl}


\bibitem[Langer et~al\mbox{.}(2021)]%
        {Langer_2021_whatdowewant}
\bibfield{author}{\bibinfo{person}{Markus Langer}, \bibinfo{person}{Daniel Oster}, \bibinfo{person}{Timo Speith}, \bibinfo{person}{Holger Hermanns}, \bibinfo{person}{Lena Kästner}, \bibinfo{person}{Eva Schmidt}, \bibinfo{person}{Andreas Sesing}, {and} \bibinfo{person}{Kevin Baum}.} \bibinfo{year}{2021}\natexlab{}.
\newblock \showarticletitle{What do we want from Explainable Artificial Intelligence (XAI)? – A stakeholder perspective on XAI and a conceptual model guiding interdisciplinary XAI research}.
\newblock \bibinfo{journal}{\emph{Artificial Intelligence}}  \bibinfo{volume}{296} (\bibinfo{date}{July} \bibinfo{year}{2021}), \bibinfo{pages}{103473}.
\newblock
\showISSN{0004-3702}
\urldef\tempurl%
\url{https://doi.org/10.1016/j.artint.2021.103473}
\showDOI{\tempurl}


\bibitem[Lee(2018)]%
        {Lee_2018}
\bibfield{author}{\bibinfo{person}{Min~Kyung Lee}.} \bibinfo{year}{2018}\natexlab{}.
\newblock \showarticletitle{Understanding perception of algorithmic decisions: Fairness, trust, and emotion in response to algorithmic management}.
\newblock \bibinfo{journal}{\emph{Big Data \& Society}} \bibinfo{volume}{5}, \bibinfo{number}{1} (\bibinfo{date}{Jan.} \bibinfo{year}{2018}), \bibinfo{pages}{205395171875668}.
\newblock
\showISSN{2053-9517}
\urldef\tempurl%
\url{https://doi.org/10.1177/2053951718756684}
\showDOI{\tempurl}


\bibitem[Lima et~al\mbox{.}(2021)]%
        {lima_human_2021}
\bibfield{author}{\bibinfo{person}{Gabriel Lima}, \bibinfo{person}{Nina Grgić-Hlača}, {and} \bibinfo{person}{Meeyoung Cha}.} \bibinfo{year}{2021}\natexlab{}.
\newblock \showarticletitle{Human {Perceptions} on {Moral} {Responsibility} of {AI}: {A} {Case} {Study} in {AI}-{Assisted} {Bail} {Decision}-{Making}}. In \bibinfo{booktitle}{\emph{Proceedings of the 2021 {CHI} {Conference} on {Human} {Factors} in {Computing} {Systems}}} \emph{(\bibinfo{series}{{CHI} '21})}. \bibinfo{publisher}{Association for Computing Machinery}, \bibinfo{address}{New York, NY, USA}, \bibinfo{pages}{1--17}.
\newblock
\showISBNx{978-1-4503-8096-6}
\urldef\tempurl%
\url{https://doi.org/10.1145/3411764.3445260}
\showDOI{\tempurl}


\bibitem[Lindlbauer et~al\mbox{.}(2019)]%
        {Lindlbauer2019}
\bibfield{author}{\bibinfo{person}{David Lindlbauer}, \bibinfo{person}{Anna~Maria Feit}, {and} \bibinfo{person}{Otmar Hilliges}.} \bibinfo{year}{2019}\natexlab{}.
\newblock \showarticletitle{Context-Aware Online Adaptation of Mixed Reality Interfaces}. In \bibinfo{booktitle}{\emph{Proceedings of the 32nd Annual ACM Symposium on User Interface Software and Technology}} (New Orleans, LA, USA) \emph{(\bibinfo{series}{UIST '19})}. \bibinfo{publisher}{Association for Computing Machinery}, \bibinfo{address}{New York, NY, USA}, \bibinfo{pages}{147–160}.
\newblock
\showISBNx{9781450368162}
\urldef\tempurl%
\url{https://doi.org/10.1145/3332165.3347945}
\showDOI{\tempurl}


\bibitem[Liu et~al\mbox{.}(2021)]%
        {liu_understanding_2021}
\bibfield{author}{\bibinfo{person}{Han Liu}, \bibinfo{person}{Vivian Lai}, {and} \bibinfo{person}{Chenhao Tan}.} \bibinfo{year}{2021}\natexlab{}.
\newblock \showarticletitle{Understanding the {Effect} of {Out}-of-distribution {Examples} and {Interactive} {Explanations} on {Human}-{AI} {Decision} {Making}}.
\newblock \bibinfo{journal}{\emph{Proceedings of the ACM on Human-Computer Interaction}} \bibinfo{volume}{5}, \bibinfo{number}{CSCW2} (\bibinfo{date}{Oct.} \bibinfo{year}{2021}), \bibinfo{pages}{1--45}.
\newblock
\showISSN{2573-0142}
\urldef\tempurl%
\url{https://doi.org/10.1145/3479552}
\showDOI{\tempurl}


\bibitem[Lu et~al\mbox{.}(2024)]%
        {lu_does_2024}
\bibfield{author}{\bibinfo{person}{Zhuoran Lu}, \bibinfo{person}{Dakuo Wang}, {and} \bibinfo{person}{Ming Yin}.} \bibinfo{year}{2024}\natexlab{}.
\newblock \showarticletitle{Does {More} {Advice} {Help}? {The} {Effects} of {Second} {Opinions} in {AI}-{Assisted} {Decision} {Making}}.
\newblock \bibinfo{journal}{\emph{Proc. ACM Hum.-Comput. Interact.}} \bibinfo{volume}{8}, \bibinfo{number}{CSCW1} (\bibinfo{date}{April} \bibinfo{year}{2024}), \bibinfo{pages}{217:1--217:31}.
\newblock
\urldef\tempurl%
\url{https://doi.org/10.1145/3653708}
\showDOI{\tempurl}


\bibitem[Ma et~al\mbox{.}(2024)]%
        {ma_are_2024}
\bibfield{author}{\bibinfo{person}{Shuai Ma}, \bibinfo{person}{Xinru Wang}, \bibinfo{person}{Ying Lei}, \bibinfo{person}{Chuhan Shi}, \bibinfo{person}{Ming Yin}, {and} \bibinfo{person}{Xiaojuan Ma}.} \bibinfo{year}{2024}\natexlab{}.
\newblock \showarticletitle{“{Are} {You} {Really} {Sure}?” {Understanding} the {Effects} of {Human} {Self}-{Confidence} {Calibration} in {AI}-{Assisted} {Decision} {Making}}. In \bibinfo{booktitle}{\emph{Proceedings of the {CHI} {Conference} on {Human} {Factors} in {Computing} {Systems}}} \emph{(\bibinfo{series}{{CHI} '24})}. \bibinfo{publisher}{Association for Computing Machinery}, \bibinfo{address}{New York, NY, USA}, \bibinfo{pages}{1--20}.
\newblock
\showISBNx{9798400703300}
\urldef\tempurl%
\url{https://doi.org/10.1145/3613904.3642671}
\showDOI{\tempurl}


\bibitem[Marcinkowski et~al\mbox{.}(2020)]%
        {Marcinkowski_2020}
\bibfield{author}{\bibinfo{person}{Frank Marcinkowski}, \bibinfo{person}{Kimon Kieslich}, \bibinfo{person}{Christopher Starke}, {and} \bibinfo{person}{Marco Lünich}.} \bibinfo{year}{2020}\natexlab{}.
\newblock \showarticletitle{Implications of AI (un-)fairness in higher education admissions: the effects of perceived AI (un-)fairness on exit, voice and organizational reputation}. In \bibinfo{booktitle}{\emph{Proceedings of the 2020 Conference on Fairness, Accountability, and Transparency}} \emph{(\bibinfo{series}{FAT* ’20})}. \bibinfo{publisher}{ACM}.
\newblock
\urldef\tempurl%
\url{https://doi.org/10.1145/3351095.3372867}
\showDOI{\tempurl}


\bibitem[Meske and Ünal(2024)]%
        {meske_investigating_2024}
\bibfield{author}{\bibinfo{person}{Christian Meske} {and} \bibinfo{person}{Erdi Ünal}.} \bibinfo{year}{2024}\natexlab{}.
\newblock \showarticletitle{Investigating the {Impact} of {Control} in {AI}-{Assisted} {Decision}-{Making} - {An} {Experimental} {Study}}. In \bibinfo{booktitle}{\emph{Proceedings of {Mensch} und {Computer} 2024}} \emph{(\bibinfo{series}{{MuC} '24})}. \bibinfo{publisher}{Association for Computing Machinery}, \bibinfo{address}{New York, NY, USA}, \bibinfo{pages}{419--423}.
\newblock
\showISBNx{9798400709982}
\urldef\tempurl%
\url{https://doi.org/10.1145/3670653.3677476}
\showDOI{\tempurl}


\bibitem[Morgeson and Campion(2003)]%
        {Morgeson_2003}
\bibfield{author}{\bibinfo{person}{Frederick~P. Morgeson} {and} \bibinfo{person}{Michael~A. Campion}.} \bibinfo{year}{2003}\natexlab{}.
\newblock \bibinfo{title}{Work Design}.
\newblock , \bibinfo{numpages}{423–452}~pages.
\newblock
\showISBNx{9780471264385}
\urldef\tempurl%
\url{https://doi.org/10.1002/0471264385.wei1217}
\showDOI{\tempurl}


\bibitem[Morgeson and Humphrey(2006)]%
        {Morgeson_WDQ_2006}
\bibfield{author}{\bibinfo{person}{Frederick~P. Morgeson} {and} \bibinfo{person}{Stephen~E. Humphrey}.} \bibinfo{year}{2006}\natexlab{}.
\newblock \showarticletitle{The Work Design Questionnaire (WDQ): Developing and validating a comprehensive measure for assessing job design and the nature of work.}
\newblock \bibinfo{journal}{\emph{Journal of Applied Psychology}} \bibinfo{volume}{91}, \bibinfo{number}{6} (\bibinfo{year}{2006}), \bibinfo{pages}{1321–1339}.
\newblock
\showISSN{0021-9010}
\urldef\tempurl%
\url{https://doi.org/10.1037/0021-9010.91.6.1321}
\showDOI{\tempurl}


\bibitem[Motowidlo et~al\mbox{.}(1986)]%
        {motowidlo_occupational_1986}
\bibfield{author}{\bibinfo{person}{Stephan~J. Motowidlo}, \bibinfo{person}{John~S. Packard}, {and} \bibinfo{person}{Michael~R. Manning}.} \bibinfo{year}{1986}\natexlab{}.
\newblock \showarticletitle{Occupational stress: {Its} causes and consequences for job performance.}
\newblock \bibinfo{journal}{\emph{Journal of Applied Psychology}} \bibinfo{volume}{71}, \bibinfo{number}{4} (\bibinfo{year}{1986}), \bibinfo{pages}{618--629}.
\newblock
\showISSN{1939-1854, 0021-9010}
\urldef\tempurl%
\url{https://doi.org/10.1037/0021-9010.71.4.618}
\showDOI{\tempurl}


\bibitem[Möhlmann et~al\mbox{.}(2021)]%
        {moehlmann_2021}
\bibfield{author}{\bibinfo{person}{Mareike Möhlmann}, \bibinfo{person}{Lior Zalmanson}, \bibinfo{person}{Ola Henfridsson}, {and} \bibinfo{person}{Robert~Wayne Gregory}.} \bibinfo{year}{2021}\natexlab{}.
\newblock \showarticletitle{Algorithmic Management of Work on Online Labor Platforms: When Matching Meets Control}.
\newblock \bibinfo{journal}{\emph{MIS Quarterly}} \bibinfo{volume}{45}, \bibinfo{number}{4} (\bibinfo{date}{Oct.} \bibinfo{year}{2021}), \bibinfo{pages}{1999–2022}.
\newblock
\showISSN{2162-9730}
\urldef\tempurl%
\url{https://doi.org/10.25300/misq/2021/15333}
\showDOI{\tempurl}


\bibitem[Nolan et~al\mbox{.}(2016)]%
        {nolan_threat_2016}
\bibfield{author}{\bibinfo{person}{Kevin Nolan}, \bibinfo{person}{Nathan Carter}, {and} \bibinfo{person}{Dev Dalal}.} \bibinfo{year}{2016}\natexlab{}.
\newblock \showarticletitle{Threat of {Technological} {Unemployment}: {Are} {Hiring} {Managers} {Discounted} for {Using} {Standardized} {Employee} {Selection} {Practices}?}
\newblock \bibinfo{journal}{\emph{Personnel Assessment and Decisions}} \bibinfo{volume}{2}, \bibinfo{number}{1} (\bibinfo{date}{July} \bibinfo{year}{2016}).
\newblock
\showISSN{2377-8822}
\urldef\tempurl%
\url{https://doi.org/10.25035/pad.2016.004}
\showDOI{\tempurl}


\bibitem[Parasuraman and Manzey(2010)]%
        {Parasuraman_2010}
\bibfield{author}{\bibinfo{person}{Raja Parasuraman} {and} \bibinfo{person}{Dietrich~H. Manzey}.} \bibinfo{year}{2010}\natexlab{}.
\newblock \showarticletitle{Complacency and Bias in Human Use of Automation: An Attentional Integration}.
\newblock \bibinfo{journal}{\emph{Human Factors: The Journal of the Human Factors and Ergonomics Society}} \bibinfo{volume}{52}, \bibinfo{number}{3} (\bibinfo{date}{June} \bibinfo{year}{2010}), \bibinfo{pages}{381–410}.
\newblock
\showISSN{1547-8181}
\urldef\tempurl%
\url{https://doi.org/10.1177/0018720810376055}
\showDOI{\tempurl}


\bibitem[Pareek et~al\mbox{.}(2024)]%
        {pareek_trust_2024}
\bibfield{author}{\bibinfo{person}{Saumya Pareek}, \bibinfo{person}{Eduardo Velloso}, {and} \bibinfo{person}{Jorge Goncalves}.} \bibinfo{year}{2024}\natexlab{}.
\newblock \showarticletitle{Trust {Development} and {Repair} in {AI}-{Assisted} {Decision}-{Making} during {Complementary} {Expertise}}. In \bibinfo{booktitle}{\emph{Proceedings of the 2024 {ACM} {Conference} on {Fairness}, {Accountability}, and {Transparency}}} \emph{(\bibinfo{series}{{FAccT} '24})}. \bibinfo{publisher}{Association for Computing Machinery}, \bibinfo{address}{New York, NY, USA}, \bibinfo{pages}{546--561}.
\newblock
\showISBNx{9798400704505}
\urldef\tempurl%
\url{https://doi.org/10.1145/3630106.3658924}
\showDOI{\tempurl}


\bibitem[Parent-Rocheleau and Parker(2022)]%
        {Parent_Rocheleau_2022}
\bibfield{author}{\bibinfo{person}{Xavier Parent-Rocheleau} {and} \bibinfo{person}{Sharon~K. Parker}.} \bibinfo{year}{2022}\natexlab{}.
\newblock \showarticletitle{Algorithms as work designers: How algorithmic management influences the design of jobs}.
\newblock \bibinfo{journal}{\emph{Human Resource Management Review}} \bibinfo{volume}{32}, \bibinfo{number}{3} (\bibinfo{date}{Sept.} \bibinfo{year}{2022}), \bibinfo{pages}{100838}.
\newblock
\showISSN{1053-4822}
\urldef\tempurl%
\url{https://doi.org/10.1016/j.hrmr.2021.100838}
\showDOI{\tempurl}


\bibitem[Parker and Hyett(2011)]%
        {parker_measurement_2011}
\bibfield{author}{\bibinfo{person}{Gordon~B. Parker} {and} \bibinfo{person}{Matthew~P. Hyett}.} \bibinfo{year}{2011}\natexlab{}.
\newblock \showarticletitle{Measurement of {Well}-{Being} in the {Workplace}: {The} {Development} of the {Work} {Well}-{Being} {Questionnaire}}.
\newblock \bibinfo{journal}{\emph{The Journal of Nervous and Mental Disease}} \bibinfo{volume}{199}, \bibinfo{number}{6} (\bibinfo{date}{June} \bibinfo{year}{2011}), \bibinfo{pages}{394}.
\newblock
\showISSN{0022-3018}
\urldef\tempurl%
\url{https://doi.org/10.1097/NMD.0b013e31821cd3b9}
\showDOI{\tempurl}


\bibitem[Parker(2014)]%
        {Parker_2014_beyondmotivation}
\bibfield{author}{\bibinfo{person}{Sharon~K. Parker}.} \bibinfo{year}{2014}\natexlab{}.
\newblock \showarticletitle{Beyond Motivation: Job and Work Design for Development, Health, Ambidexterity, and More}.
\newblock \bibinfo{journal}{\emph{Annual Review of Psychology}} \bibinfo{volume}{65}, \bibinfo{number}{1} (\bibinfo{date}{Jan.} \bibinfo{year}{2014}), \bibinfo{pages}{661–691}.
\newblock
\showISSN{1545-2085}
\urldef\tempurl%
\url{https://doi.org/10.1146/annurev-psych-010213-115208}
\showDOI{\tempurl}


\bibitem[Parker and Grote(2020)]%
        {Parker_2020_algorithms}
\bibfield{author}{\bibinfo{person}{Sharon~K. Parker} {and} \bibinfo{person}{Gudela Grote}.} \bibinfo{year}{2020}\natexlab{}.
\newblock \showarticletitle{Automation, Algorithms, and Beyond: Why Work Design Matters More Than Ever in a Digital World}.
\newblock \bibinfo{journal}{\emph{Applied Psychology}} \bibinfo{volume}{71}, \bibinfo{number}{4} (\bibinfo{date}{Feb.} \bibinfo{year}{2020}), \bibinfo{pages}{1171–1204}.
\newblock
\showISSN{1464-0597}
\urldef\tempurl%
\url{https://doi.org/10.1111/apps.12241}
\showDOI{\tempurl}


\bibitem[Parker and Knight(2023)]%
        {Parker_SMART_2023}
\bibfield{author}{\bibinfo{person}{Sharon~K. Parker} {and} \bibinfo{person}{Caroline Knight}.} \bibinfo{year}{2023}\natexlab{}.
\newblock \showarticletitle{The <scp>SMART</scp> model of work design: A higher order structure to help see the wood from the trees}.
\newblock \bibinfo{journal}{\emph{Human Resource Management}} \bibinfo{volume}{63}, \bibinfo{number}{2} (\bibinfo{date}{Nov.} \bibinfo{year}{2023}), \bibinfo{pages}{265–291}.
\newblock
\showISSN{1099-050X}
\urldef\tempurl%
\url{https://doi.org/10.1002/hrm.22200}
\showDOI{\tempurl}


\bibitem[Passalacqua et~al\mbox{.}(2024)]%
        {passalacqua_practice_2024}
\bibfield{author}{\bibinfo{person}{Mario Passalacqua}, \bibinfo{person}{Robert Pellerin}, \bibinfo{person}{Esma Yahia}, \bibinfo{person}{Florian Magnani}, \bibinfo{person}{Frédéric Rosin}, \bibinfo{person}{Laurent Joblot}, {and} \bibinfo{person}{Pierre-Majorique Léger}.} \bibinfo{year}{2024}\natexlab{}.
\newblock \showarticletitle{Practice {With} {Less} {AI} {Makes} {Perfect}: {Partially} {Automated} {AI} {During} {Training} {Leads} to {Better} {Worker} {Motivation}, {Engagement}, and {Skill} {Acquisition}}.
\newblock \bibinfo{journal}{\emph{International Journal of Human–Computer Interaction}} \bibinfo{volume}{0}, \bibinfo{number}{0} (\bibinfo{date}{March} \bibinfo{year}{2024}), \bibinfo{pages}{1--21}.
\newblock
\showISSN{1044-7318}
\urldef\tempurl%
\url{https://doi.org/10.1080/10447318.2024.2319914}
\showDOI{\tempurl}
\newblock
\shownote{Publisher: Taylor \& Francis \_eprint: https://doi.org/10.1080/10447318.2024.2319914}.


\bibitem[Poursabzi-Sangdeh et~al\mbox{.}(2021)]%
        {poursabzi-sangdeh_manipulating_2021}
\bibfield{author}{\bibinfo{person}{Forough Poursabzi-Sangdeh}, \bibinfo{person}{Daniel~G. Goldstein}, \bibinfo{person}{Jake~M. Hofman}, \bibinfo{person}{Jennifer~Wortman Vaughan}, {and} \bibinfo{person}{Hanna Wallach}.} \bibinfo{year}{2021}\natexlab{}.
\newblock \bibinfo{title}{Manipulating and {Measuring} {Model} {Interpretability}}.
\newblock
\newblock
\urldef\tempurl%
\url{http://arxiv.org/abs/1802.07810}
\showURL{%
\tempurl}
\newblock
\shownote{arXiv:1802.07810 [cs]}.


\bibitem[Rastogi et~al\mbox{.}(2022)]%
        {rastogi_deciding_2022}
\bibfield{author}{\bibinfo{person}{Charvi Rastogi}, \bibinfo{person}{Yunfeng Zhang}, \bibinfo{person}{Dennis Wei}, \bibinfo{person}{Kush~R. Varshney}, \bibinfo{person}{Amit Dhurandhar}, {and} \bibinfo{person}{Richard Tomsett}.} \bibinfo{year}{2022}\natexlab{}.
\newblock \showarticletitle{Deciding {Fast} and {Slow}: {The} {Role} of {Cognitive} {Biases} in {AI}-assisted {Decision}-making}.
\newblock \bibinfo{journal}{\emph{Proc. ACM Hum.-Comput. Interact.}} \bibinfo{volume}{6}, \bibinfo{number}{CSCW1} (\bibinfo{date}{April} \bibinfo{year}{2022}), \bibinfo{pages}{83:1--83:22}.
\newblock
\urldef\tempurl%
\url{https://doi.org/10.1145/3512930}
\showDOI{\tempurl}


\bibitem[Robbemond et~al\mbox{.}(2022)]%
        {robbemond_understanding_2022}
\bibfield{author}{\bibinfo{person}{Vincent Robbemond}, \bibinfo{person}{Oana Inel}, {and} \bibinfo{person}{Ujwal Gadiraju}.} \bibinfo{year}{2022}\natexlab{}.
\newblock \showarticletitle{Understanding the {Role} of {Explanation} {Modality} in {AI}-assisted {Decision}-making}. In \bibinfo{booktitle}{\emph{Proceedings of the 30th {ACM} {Conference} on {User} {Modeling}, {Adaptation} and {Personalization}}} \emph{(\bibinfo{series}{{UMAP} '22})}. \bibinfo{publisher}{Association for Computing Machinery}, \bibinfo{address}{New York, NY, USA}, \bibinfo{pages}{223--233}.
\newblock
\showISBNx{978-1-4503-9207-5}
\urldef\tempurl%
\url{https://doi.org/10.1145/3503252.3531311}
\showDOI{\tempurl}


\bibitem[Ryan and Deci(2000)]%
        {Ryan_Deci_SDT_2000}
\bibfield{author}{\bibinfo{person}{Richard~M. Ryan} {and} \bibinfo{person}{Edward~L. Deci}.} \bibinfo{year}{2000}\natexlab{}.
\newblock \showarticletitle{Self-determination theory and the facilitation of intrinsic motivation, social development, and well-being.}
\newblock \bibinfo{journal}{\emph{American Psychologist}} \bibinfo{volume}{55}, \bibinfo{number}{1} (\bibinfo{year}{2000}), \bibinfo{pages}{68–78}.
\newblock
\showISSN{0003-066X}
\urldef\tempurl%
\url{https://doi.org/10.1037//0003-066x.55.1.68}
\showDOI{\tempurl}


\bibitem[Ryan and Deci(2017)]%
        {ryan_deci_SDT_2017}
\bibfield{author}{\bibinfo{person}{Richard~M. Ryan} {and} \bibinfo{person}{Edward~L. Deci}.} \bibinfo{year}{2017}\natexlab{}.
\newblock \bibinfo{booktitle}{\emph{Self-Determination Theory: Basic Psychological Needs in Motivation, Development, and Wellness}}.
\newblock \bibinfo{publisher}{Guilford Press}.
\newblock
\showISBNx{9781462538966}
\urldef\tempurl%
\url{https://doi.org/10.1521/978.14625/28806}
\showDOI{\tempurl}


\bibitem[Schoeffer et~al\mbox{.}(2024)]%
        {Schoeffer_2024}
\bibfield{author}{\bibinfo{person}{Jakob Schoeffer}, \bibinfo{person}{Maria De-Arteaga}, {and} \bibinfo{person}{Niklas Kühl}.} \bibinfo{year}{2024}\natexlab{}.
\newblock \showarticletitle{Explanations, Fairness, and Appropriate Reliance in Human-AI Decision-Making}. In \bibinfo{booktitle}{\emph{Proceedings of the CHI Conference on Human Factors in Computing Systems}} \emph{(\bibinfo{series}{CHI ’24})}. \bibinfo{publisher}{ACM}, \bibinfo{pages}{1–18}.
\newblock
\urldef\tempurl%
\url{https://doi.org/10.1145/3613904.3642621}
\showDOI{\tempurl}


\bibitem[Shah et~al\mbox{.}(2018)]%
        {shah2018airsim}
\bibfield{author}{\bibinfo{person}{Shital Shah}, \bibinfo{person}{Debadeepta Dey}, \bibinfo{person}{Chris Lovett}, {and} \bibinfo{person}{Ashish Kapoor}.} \bibinfo{year}{2018}\natexlab{}.
\newblock \showarticletitle{Airsim: High-fidelity visual and physical simulation for autonomous vehicles}. In \bibinfo{booktitle}{\emph{Field and Service Robotics: Results of the 11th International Conference}}. Springer, \bibinfo{pages}{621--635}.
\newblock


\bibitem[Spreitzer(1995)]%
        {spreitzer_empirical_1995}
\bibfield{author}{\bibinfo{person}{Gretchen~M. Spreitzer}.} \bibinfo{year}{1995}\natexlab{}.
\newblock \showarticletitle{An empirical test of a comprehensive model of intrapersonal empowerment in the workplace}.
\newblock \bibinfo{journal}{\emph{American Journal of Community Psychology}} \bibinfo{volume}{23}, \bibinfo{number}{5} (\bibinfo{date}{Oct.} \bibinfo{year}{1995}), \bibinfo{pages}{601--629}.
\newblock
\showISSN{0091-0562, 1573-2770}
\urldef\tempurl%
\url{https://doi.org/10.1007/BF02506984}
\showDOI{\tempurl}


\bibitem[Stanley et~al\mbox{.}(2021)]%
        {Stanley_meta_2021}
\bibfield{author}{\bibinfo{person}{Peter~J. Stanley}, \bibinfo{person}{Nicola~S. Schutte}, {and} \bibinfo{person}{Wendy~J. Phillips}.} \bibinfo{year}{2021}\natexlab{}.
\newblock \showarticletitle{A meta-analytic investigation of the relationship between basic psychological need satisfaction and affect}.
\newblock \bibinfo{journal}{\emph{Journal of Positive School Psychology}} \bibinfo{volume}{5}, \bibinfo{number}{1} (\bibinfo{date}{April} \bibinfo{year}{2021}), \bibinfo{pages}{1–16}.
\newblock
\showISSN{2717-7564}
\urldef\tempurl%
\url{https://doi.org/10.47602/jpsp.v5i1.210}
\showDOI{\tempurl}


\bibitem[Sterz et~al\mbox{.}(2024)]%
        {Sterz_2024}
\bibfield{author}{\bibinfo{person}{Sarah Sterz}, \bibinfo{person}{Kevin Baum}, \bibinfo{person}{Sebastian Biewer}, \bibinfo{person}{Holger Hermanns}, \bibinfo{person}{Anne Lauber-Rönsberg}, \bibinfo{person}{Philip Meinel}, {and} \bibinfo{person}{Markus Langer}.} \bibinfo{year}{2024}\natexlab{}.
\newblock \showarticletitle{On the Quest for Effectiveness in Human Oversight: Interdisciplinary Perspectives}. In \bibinfo{booktitle}{\emph{The 2024 ACM Conference on Fairness, Accountability, and Transparency}} \emph{(\bibinfo{series}{FAccT ’24}, Vol.~\bibinfo{volume}{15})}. \bibinfo{publisher}{ACM}, \bibinfo{pages}{2495–2507}.
\newblock
\urldef\tempurl%
\url{https://doi.org/10.1145/3630106.3659051}
\showDOI{\tempurl}


\bibitem[Steyvers and Kumar(2023)]%
        {steyvers_three_nodate}
\bibfield{author}{\bibinfo{person}{Mark Steyvers} {and} \bibinfo{person}{Aakriti Kumar}.} \bibinfo{year}{2023}\natexlab{}.
\newblock \showarticletitle{Three Challenges for AI-Assisted Decision-Making}.
\newblock \bibinfo{journal}{\emph{Perspectives on Psychological Science}} (\bibinfo{year}{2023}), \bibinfo{pages}{17456916231181102}.
\newblock
\urldef\tempurl%
\url{https://doi.org/10.1177/17456916231181102}
\showDOI{\tempurl}
\showeprint{https://doi.org/10.1177/17456916231181102}
\newblock
\shownote{PMID: 37439761}.


\bibitem[Stiglbauer and Kovacs(2018)]%
        {Stiglbauer_2018}
\bibfield{author}{\bibinfo{person}{Barbara Stiglbauer} {and} \bibinfo{person}{Carrie Kovacs}.} \bibinfo{year}{2018}\natexlab{}.
\newblock \showarticletitle{The more, the better? Curvilinear effects of job autonomy on well-being from vitamin model and PE-fit theory perspectives.}
\newblock \bibinfo{journal}{\emph{Journal of Occupational Health Psychology}} \bibinfo{volume}{23}, \bibinfo{number}{4} (\bibinfo{date}{Oct.} \bibinfo{year}{2018}), \bibinfo{pages}{520–536}.
\newblock
\showISSN{1076-8998}
\urldef\tempurl%
\url{https://doi.org/10.1037/ocp0000107}
\showDOI{\tempurl}


\bibitem[Strich et~al\mbox{.}(2021)]%
        {Strich_2021}
\bibfield{author}{\bibinfo{person}{Franz Strich}, \bibinfo{person}{Anne-Sophie Mayer}, {and} \bibinfo{person}{Marina Fiedler}.} \bibinfo{year}{2021}\natexlab{}.
\newblock \showarticletitle{What Do I Do in a World of Artificial Intelligence? Investigating the Impact of Substitutive Decision-Making AI Systems on Employees’ Professional Role Identity}.
\newblock \bibinfo{journal}{\emph{Journal of the Association for Information Systems}} \bibinfo{volume}{22}, \bibinfo{number}{2} (\bibinfo{year}{2021}), \bibinfo{pages}{304–324}.
\newblock
\showISSN{1536-9323}
\urldef\tempurl%
\url{https://doi.org/10.17705/1jais.00663}
\showDOI{\tempurl}


\bibitem[Swaroop et~al\mbox{.}(2024)]%
        {swaroop_accuracy-time_2024}
\bibfield{author}{\bibinfo{person}{Siddharth Swaroop}, \bibinfo{person}{Zana Buçinca}, \bibinfo{person}{Krzysztof~Z. Gajos}, {and} \bibinfo{person}{Finale Doshi-Velez}.} \bibinfo{year}{2024}\natexlab{}.
\newblock \showarticletitle{Accuracy-{Time} {Tradeoffs} in {AI}-{Assisted} {Decision} {Making} under {Time} {Pressure}}. In \bibinfo{booktitle}{\emph{Proceedings of the 29th {International} {Conference} on {Intelligent} {User} {Interfaces}}} \emph{(\bibinfo{series}{{IUI} '24})}. \bibinfo{publisher}{Association for Computing Machinery}, \bibinfo{address}{New York, NY, USA}, \bibinfo{pages}{138--154}.
\newblock
\showISBNx{9798400705083}
\urldef\tempurl%
\url{https://doi.org/10.1145/3640543.3645206}
\showDOI{\tempurl}


\bibitem[Tang et~al\mbox{.}(2019)]%
        {Tang_review_2019}
\bibfield{author}{\bibinfo{person}{Minmin Tang}, \bibinfo{person}{Dahua Wang}, {and} \bibinfo{person}{Alain Guerrien}.} \bibinfo{year}{2019}\natexlab{}.
\newblock \showarticletitle{A systematic review and meta‐analysis on basic psychological need satisfaction, motivation, and well‐being in later life: Contributions of self‐determination theory}.
\newblock \bibinfo{journal}{\emph{PsyCh Journal}} \bibinfo{volume}{9}, \bibinfo{number}{1} (\bibinfo{date}{June} \bibinfo{year}{2019}), \bibinfo{pages}{5–33}.
\newblock
\showISSN{2046-0260}
\urldef\tempurl%
\url{https://doi.org/10.1002/pchj.293}
\showDOI{\tempurl}


\bibitem[Tay et~al\mbox{.}(2023)]%
        {Tay_2023}
\bibfield{author}{\bibinfo{person}{Louis Tay}, \bibinfo{person}{Cassondra Batz-Barbarich}, \bibinfo{person}{Liu-Qin Yang}, {and} \bibinfo{person}{Christopher~W Wiese}.} \bibinfo{year}{2023}\natexlab{}.
\newblock \showarticletitle{Well-Being: The Ultimate Criterion for Organizational Sciences}.
\newblock  (\bibinfo{date}{Aug.} \bibinfo{year}{2023}).
\newblock
\urldef\tempurl%
\url{https://doi.org/10.31234/osf.io/4v2y5}
\showDOI{\tempurl}


\bibitem[Ulfert et~al\mbox{.}(2022)]%
        {ulfert_agent_autonomy}
\bibfield{author}{\bibinfo{person}{Anna-Sophie Ulfert}, \bibinfo{person}{Conny~H. Antoni}, {and} \bibinfo{person}{Thomas Ellwart}.} \bibinfo{year}{2022}\natexlab{}.
\newblock \showarticletitle{The role of agent autonomy in using decision support systems at work}.
\newblock \bibinfo{journal}{\emph{Computers in Human Behavior}}  \bibinfo{volume}{126} (\bibinfo{year}{2022}), \bibinfo{pages}{106987}.
\newblock
\showISSN{0747-5632}
\urldef\tempurl%
\url{https://doi.org/10.1016/j.chb.2021.106987}
\showDOI{\tempurl}


\bibitem[Van~den Broeck et~al\mbox{.}(2016)]%
        {Van_den_Broeck_Review_2016}
\bibfield{author}{\bibinfo{person}{Anja Van~den Broeck}, \bibinfo{person}{D.~Lance Ferris}, \bibinfo{person}{Chu-Hsiang Chang}, {and} \bibinfo{person}{Christopher~C. Rosen}.} \bibinfo{year}{2016}\natexlab{}.
\newblock \showarticletitle{A Review of Self-Determination Theory’s Basic Psychological Needs at Work}.
\newblock \bibinfo{journal}{\emph{Journal of Management}} \bibinfo{volume}{42}, \bibinfo{number}{5} (\bibinfo{date}{March} \bibinfo{year}{2016}), \bibinfo{pages}{1195–1229}.
\newblock
\showISSN{1557-1211}
\urldef\tempurl%
\url{https://doi.org/10.1177/0149206316632058}
\showDOI{\tempurl}


\bibitem[Van~den Broeck et~al\mbox{.}(2010)]%
        {Van_den_Broeck_Capturing_2010}
\bibfield{author}{\bibinfo{person}{Anja Van~den Broeck}, \bibinfo{person}{Maarten Vansteenkiste}, \bibinfo{person}{Hans De~Witte}, \bibinfo{person}{Bart Soenens}, {and} \bibinfo{person}{Willy Lens}.} \bibinfo{year}{2010}\natexlab{}.
\newblock \showarticletitle{Capturing autonomy, competence, and relatedness at work: Construction and initial validation of the Work‐related Basic Need Satisfaction scale}.
\newblock \bibinfo{journal}{\emph{Journal of Occupational and Organizational Psychology}} \bibinfo{volume}{83}, \bibinfo{number}{4} (\bibinfo{date}{Dec.} \bibinfo{year}{2010}), \bibinfo{pages}{981–1002}.
\newblock
\showISSN{2044-8325}
\urldef\tempurl%
\url{https://doi.org/10.1348/096317909x481382}
\showDOI{\tempurl}


\bibitem[Van~Dick et~al\mbox{.}(2001)]%
        {van_dick_job_2001}
\bibfield{author}{\bibinfo{person}{Rolf Van~Dick}, \bibinfo{person}{Christiane Schnitger}, \bibinfo{person}{Carla Schwartzmann-Buchelt}, {and} \bibinfo{person}{Ulrich Wagner}.} \bibinfo{year}{2001}\natexlab{}.
\newblock \showarticletitle{Der {Job} {Diagnostic} {Survey} im {Bildungsbereich}: {Eine} Überprüfung der {Gültigkeit} des {Job} {Characteristics} {Model} bei {Lehrerinnen} und {Lehrern}, {Hochschulangehörigen} und {Erzieherinnen} mit berufsspezifischen {Weiterentwicklungen} des {JDS}}.
\newblock \bibinfo{journal}{\emph{Zeitschrift für Arbeits- und Organisationspsychologie A\&O}} \bibinfo{volume}{45}, \bibinfo{number}{2} (\bibinfo{date}{April} \bibinfo{year}{2001}), \bibinfo{pages}{74--92}.
\newblock
\showISSN{0932-4089, 2190-6270}
\urldef\tempurl%
\url{https://doi.org/10.1026//0932-4089.45.2.74}
\showDOI{\tempurl}


\bibitem[Vollmeyer and Rheinberg(2000)]%
        {Vollmeyer_motivation_2000}
\bibfield{author}{\bibinfo{person}{Regina Vollmeyer} {and} \bibinfo{person}{Falko Rheinberg}.} \bibinfo{year}{2000}\natexlab{}.
\newblock \showarticletitle{Does motivation affect performance via persistence?}
\newblock \bibinfo{journal}{\emph{Learning and Instruction}} \bibinfo{volume}{10}, \bibinfo{number}{4} (\bibinfo{date}{Aug.} \bibinfo{year}{2000}), \bibinfo{pages}{293–309}.
\newblock
\showISSN{0959-4752}
\urldef\tempurl%
\url{https://doi.org/10.1016/s0959-4752(99)00031-6}
\showDOI{\tempurl}


\bibitem[Wang and Yin(2021)]%
        {wang_are_2021}
\bibfield{author}{\bibinfo{person}{Xinru Wang} {and} \bibinfo{person}{Ming Yin}.} \bibinfo{year}{2021}\natexlab{}.
\newblock \showarticletitle{Are {Explanations} {Helpful}? {A} {Comparative} {Study} of the {Effects} of {Explanations} in {AI}-{Assisted} {Decision}-{Making}}. In \bibinfo{booktitle}{\emph{26th {International} {Conference} on {Intelligent} {User} {Interfaces}}}. \bibinfo{publisher}{ACM}, \bibinfo{address}{College Station TX USA}, \bibinfo{pages}{318--328}.
\newblock
\showISBNx{978-1-4503-8017-1}
\urldef\tempurl%
\url{https://doi.org/10.1145/3397481.3450650}
\showDOI{\tempurl}


\bibitem[Wang and Yin(2023)]%
        {wang_watch_2023}
\bibfield{author}{\bibinfo{person}{Xinru Wang} {and} \bibinfo{person}{Ming Yin}.} \bibinfo{year}{2023}\natexlab{}.
\newblock \showarticletitle{Watch {Out} for {Updates}: {Understanding} the {Effects} of {Model} {Explanation} {Updates} in {AI}-{Assisted} {Decision} {Making}}. In \bibinfo{booktitle}{\emph{Proceedings of the 2023 {CHI} {Conference} on {Human} {Factors} in {Computing} {Systems}}} \emph{(\bibinfo{series}{{CHI} '23})}. \bibinfo{publisher}{Association for Computing Machinery}, \bibinfo{address}{New York, NY, USA}, \bibinfo{pages}{1--19}.
\newblock
\showISBNx{978-1-4503-9421-5}
\urldef\tempurl%
\url{https://doi.org/10.1145/3544548.3581366}
\showDOI{\tempurl}


\bibitem[Yao and Siegel(2021)]%
        {yao_influence_2021}
\bibfield{author}{\bibinfo{person}{Elvin Yao} {and} \bibinfo{person}{Jason~T. Siegel}.} \bibinfo{year}{2021}\natexlab{}.
\newblock \showarticletitle{The influence of perceptions of intentionality and controllability on perceived responsibility: {Applying} attribution theory to people’s responses to social transgression in the {COVID}-19 pandemic.}
\newblock \bibinfo{journal}{\emph{Motivation Science}} \bibinfo{volume}{7}, \bibinfo{number}{2} (\bibinfo{date}{June} \bibinfo{year}{2021}), \bibinfo{pages}{199--206}.
\newblock
\showISSN{2333-8113}
\urldef\tempurl%
\url{https://doi.org/10.1037/mot0000220}
\showDOI{\tempurl}
\newblock
\shownote{Publisher: Educational Publishing Foundation}.


\bibitem[Zerilli et~al\mbox{.}(2019)]%
        {Zerilli_2019}
\bibfield{author}{\bibinfo{person}{John Zerilli}, \bibinfo{person}{Alistair Knott}, \bibinfo{person}{James Maclaurin}, {and} \bibinfo{person}{Colin Gavaghan}.} \bibinfo{year}{2019}\natexlab{}.
\newblock \showarticletitle{Algorithmic Decision-Making and the Control Problem}.
\newblock \bibinfo{journal}{\emph{Minds and Machines}} \bibinfo{volume}{29}, \bibinfo{number}{4} (\bibinfo{date}{Dec.} \bibinfo{year}{2019}), \bibinfo{pages}{555–578}.
\newblock
\showISSN{1572-8641}
\urldef\tempurl%
\url{https://doi.org/10.1007/s11023-019-09513-7}
\showDOI{\tempurl}


\bibitem[Zhang et~al\mbox{.}(2020)]%
        {zhang_effect_2020}
\bibfield{author}{\bibinfo{person}{Yunfeng Zhang}, \bibinfo{person}{Q.~Vera Liao}, {and} \bibinfo{person}{Rachel K.~E. Bellamy}.} \bibinfo{year}{2020}\natexlab{}.
\newblock \showarticletitle{Effect of confidence and explanation on accuracy and trust calibration in {AI}-assisted decision making}. In \bibinfo{booktitle}{\emph{Proceedings of the 2020 {Conference} on {Fairness}, {Accountability}, and {Transparency}}}. \bibinfo{publisher}{ACM}, \bibinfo{address}{Barcelona Spain}, \bibinfo{pages}{295--305}.
\newblock
\showISBNx{978-1-4503-6936-7}
\urldef\tempurl%
\url{https://doi.org/10.1145/3351095.3372852}
\showDOI{\tempurl}


\end{thebibliography}

\appendix
\newpage
\section{Additional Questionnaire Items}
\begin{table*}[h]
    \centering
\caption{Exploratory questionnaire items and measured concepts}
\label{tab:additional_questionnaire_items}
    \begin{tabular}{>{\raggedright\arraybackslash}p{0.3\linewidth} >{\raggedright\arraybackslash}p{0.6\linewidth} c} \toprule
        Concept & Question(s)&Ref\\ \midrule
        Personal Responsibility& 'I was responsible for the crash of the drone.' & \cite{hinds_whose_2004}\\ 
        Causality& 'I caused the drone to crash.' & \cite{gailey_attribution_2008}\\
        &'I could have avoided crashing the drone.' & \\
        &'I could have prevented the drone from crashing.' & \\
        Knowledge&'I could have foreseen the crash of the drone.'  &\cite{gailey_attribution_2008}\\
        &'I understand why the drone crashed.' & \\
        Intention&'I had the intention of crashing the drone.' &\cite{gailey_attribution_2008}\\
        Control&'I felt like I was in control over the crash of the drone.'   & \cite{nolan_threat_2016}\\
        Trust&'I can trust the system.'  & \cite{korber_theoretical_2019}\\\
        &'I can rely on the available options.' &\\
        Disengagement&'I feel that I have become disconnected from my work task. ' &\cite{demerouti2008oldenburg}\\
        Exhaustion&'I feel tired performing the task. ' &\cite{demerouti2008oldenburg}\\
        Stress&'I am currently stressed by the task.'   & \cite{motowidlo_occupational_1986} \\
        Blame&'To what extent do you think you are to blame for the drone crash?   & \cite{ames_intentional_2013} \\
        &'To what extent do you think the system is to blame for the drone crash?' & \\
        &'To what extent do you think you should be  morally condemned for the crash of the drone?' & \\
        Punishment&'To what extent should you be punished for the crash of the drone?'   & \cite{yao_influence_2021}\\
        &'To what extent should the provider of the system be punished for the crash of the drone?'  & \\
        Performance&'I was successful in accomplishing the task.'   & \cite{hart_development_1988}\\
        Physical Demand&'The task was physically demanding.' & \\
        Effort&'I had to work hard to accomplish my level of performance.'  & \\
        Frustration&'I was irritated, stressed, and annoyed during the task.'   &\\\bottomrule
    \end{tabular}
\end{table*}

\end{document}